\documentclass[11pt]{article}
\pdfoutput=1
\usepackage{ifpdf}
\usepackage{jcappub}\usepackage[compatibility=false]{caption} % for details on the use of the package, please
\usepackage{tikz}
\usepackage[T1]{fontenc} % if needed
\usepackage{tabularx}
\usepackage{natbib}
\usepackage{ae,aecompl,bm}
\usepackage{xcolor}
\usepackage{subcaption}

\usepackage{natbib}
\usepackage{amsbsy}
\usepackage{amssymb}
\usepackage{amsmath}
\usepackage{graphicx}
\usepackage[justification=centering]{caption}
\usepackage{mathrsfs}
\usepackage{array}
\usepackage{mathrsfs}% for script P
\usepackage{caption}
\usepackage{xfrac}
\usepackage{booktabs}
\usepackage{amsthm}
\usepackage{hyperref}
\hypersetup{colorlinks=true, citecolor=blue, urlcolor=blue, linkcolor=blue}

\usepackage{mathtools} % Need for cases.
\usepackage{graphicx}% Include figure files
\usepackage{dcolumn}% Align table columns on decimal point
\usepackage{bm}% bold math
\usepackage{etoolbox}
\apptocmd{\sloppy}{\hbadness 10000\relax}{}{}

\usepackage{xcolor}
\usepackage{amsthm}
\usepackage{pifont}

\usepackage [english]{babel}
\usepackage [autostyle, english = american]{csquotes}
\usepackage{natbib}

\usetikzlibrary{positioning}

\date{}
\def\be{\begin{equation}}
\def\ee{\end{equation}}

\title{ A semi-cosmographic approach to study cosmological evolution in phase space}% Force line breaks with \\
\author[a]{Pankaj Chavan,}
\author[a,1]{Tapomoy Guha Sarkar, \note{Corresponding author.}}
\author[b]{Chandrachud B. V. Dash}
\author[c] {and Anjan A Sen}

\affiliation[a]{Department of Physics, Birla Institute of Technology and Science, Pilani, \\Rajasthan, India.}
\affiliation[b]{Astrophysics Research Centre \& School of Mathematics, Statistics and Computer Science, University of KwaZulu-Natal, Durban, South Africa.}
\affiliation[c] {Centre for Theoretical Physics, Jamia Millia Islamia, New Delhi-110025, India.}

\emailAdd{chavanpankaj09@gmail.com}
\emailAdd{cb.vaswar@gmail.com}
\emailAdd{tapomoy@pilani.bits-pilani.ac.in}
\emailAdd{aasen@jmi.ac.in}
\abstract{
The signature of Baryon Acoustic Oscillation in the clustering of dark-matter tracers allows us to measure $(D_A(z), H(z))$ independently. Treating these as conjugate variables, we are motivated to study cosmological evolution in the phase space of dimensionless variables $x = H_0 D_A/c$ and $p = dx/dz$.  The dynamical variables $(x(z),p(z))$ can be integrated for a known set of equation of state parameters for different matter/energy components. However, to avoid any preference for specific dark energy models, we adopt a cosmographic approach. We consider two scenarios where the Luminosity distance is expanded as Pad\'e rational approximants using expansion in terms of $z$ and $(1+z)^{1/2}$ respectively. However, instead of  directly using the Pad\'e ratios to fit  kinematic quantities with data, we adopt an alternative  approach where the evolution of the cold dark matter sector is incorporated in our analysis through a semi-cosmographic equation of state, which is then, used to solve the dynamical problem in the phase space. The semi-cosmographic $(D_A(z), H(z))$, thus obtained, is fitted with BAO data from DESI DR1, cosmic chronometer (CC) data and SNIa data from Pantheon+ respectively. We also consider a futuristic 21-cm intensity mapping experiment for error projections. We further use the semi-cosmographic fitting to reconstruct some diagnostics of background cosmology and compare our results for the two scenarios of Pad\'e expansions. }

\begin{document}
\maketitle
\flushbottom
\section{Introduction}
There is compelling observational evidence for the accelerated expansion of the Universe from a host of cosmological probes \cite{Perlmutter_1997, Spergel_2003, Hinshaw_2003, scranton2003physical, Eisenstein_2005, McDonald_2007, Riess_2016}. However, a theoretical understanding of dark energy DE \cite{Ratra-Peebles_1988, Padmanabhan_2003, amendola_tsujikawa_2010, bamba2012dark} as a potential cause for this cosmic acceleration remains largely uncertain even today. 
Although, the widely recognized framework of the $\Lambda$CDM model \cite{Ratra-Peebles_1988, carroll2001cosmological, Bulletal_2016} provides the broad cosmological paradigm,  a closer scrutiny indicates theoretical challenges \cite{weinberg1989cosmological, PhysRevLett.82.896,  Padmanabhan_2003} and tension with observed data \cite{copeland2006dynamics, burgess2015cosmological, Anchordoqui_2021, di2021realm, abdalla2022cosmology, perivolaropoulos2022challenges}. Notable here is the Hubble-tension which indicates that the value of  $H_0$ measured implicitly from high redshift CMBR observation \cite{kamionkowski2022hubbletensionearlydark, PhysRevD.104.083509,dutcher2021measurements,jain2003cross,huterer2002weak,amendola2008measuring,martinet2021probing}, Baryon Acoustic Oscillation (BAO) signature (in galaxy clustering \cite{seo2007improved}  or in the Ly-$\alpha$ forest \cite{Slosar_2009, baolyman2020}), Big Bang Nucleosynthesis (BBN) \cite{schoneberg_2019_BBN} and Supernova (SNIa) observations \cite{sandage2006hubble, riess2021cosmic, beaton2016carnegie, freedman2020calibration,blakeslee2021hubble},  consistently disagrees ($> 4\sigma)$ with direct  low redshift estimates from  distance measurements with HST \cite{hst-hubble}  or Cepheids (SH0ES) \cite{riess2022comprehensive}. Confronting the observational challenges \cite{di2021snowmass2021,wong2020h0licow,abbott2022dark,verde2013planck,fields2011primordial} faced by the  standard $\Lambda$CDM cosmological model, is the vast theoretical landscape of diverse DE models \cite{di2021realm}. These alternative models attempt to address the issues either by introducing additional terms in the matter sector, often involving scalar fields with new couplings \cite{Ratra-Peebles_1988, Steinhardt_1998, PhysRevLett.82.896, scherrer2008thawing}, or by modifying Einstein's theory of general relativity \cite{amendola_tsujikawa_2010, Khoury_2004, Starobinsky_2007, Hu_2007, nojiri2007introduction}.

In the absence of a satisfactory theory that is consistent with all the observations,  there is an emerging stress on data-driven model-independent frameworks.
This approach is facilitated by several independent cosmological missions that 
measure the expansion history over a wide redshift range and with an ever-improving level of precision.
The extreme examples of complete model agnosticism are the Machine Learning \cite{Rasmussen_2005_GP_ML} based approaches 
like the Gaussian process reconstruction \cite{Holsclaw_2011_GPR,Shafieloo_2012_GPR, Jesus_2024_GPR, Dinda2024_GP_cosmography,Velazquez_mukherjee_GPR_2024, Dinda_2025, Purba_Mukherjee-Anjan_sen_GPR_2024}. These data driven approaches,
are quite agnostic about the form of the function that is being reconstructed. This allows for the possibility of alleviating degeneracies between different theoretical models. However, there are still some weak inductive biases related to the specific shape of the kernel used, that can effectively impose mild assumptions.

While this approach is aligned to the pure empirical nature of scientific inquiry, and privileges observations over theory, it disallows the inclusion of any theoretical understanding of the evolution dynamics based on known physical laws or physical intuition.

Another commonly adopted model-independent approach is \textit{cosmography} \cite{Weinberg_1972_cosmography, Sahni2003_statefinder, Visser_2004_jerk_snap, Bonici_2019_Dynamical_DE, bolotin_2018_applied_cosmography}.
This approach aims to shift the focus of attention from any assumption about the fundamental underlying dynamics and,  instead attempts to constrain the kinematics of the Universe.
Observable quantities such as the cosmological distances or the
Hubble parameter are expanded as a power series in redshift, with the expansion coefficients related to various kimematic quantities \cite{Visser_2015_cosmography, Dunsby_Luongo_2016_cosmography, Capozziello_2019_cosmography, Busti_2015_cosmography, Visser2005-cosmography, Yang_Aritra_banerjee_2020_cosmography, Aviles_2013_cosmography, Aviles_2012_cosmography_y_variable, Aviles_bravetti_cosmography_2013, AVILES_2017_cosmography}. These kinematic quantities, which involve the derivatives of the scale factor, are then constrained using the observed data.
The key problem in a cosmographic approach is that the series expansion diverges for $z\geq 1$ \cite{Cattoen_2007_convergence, Capozziello_2019_cosmography, Capozziello_2020_cosmography, Gruber_Luongo_2014_cosmography, Lobo_2020}.
In such situations, keeping more terms in the expansion does not yield anything meaningful, since the radius of convergence is small. This makes the method devoid of much predictive power at high redshifts.
This is particularly concerning, since most of the recent Supernovae and BAO data are at high redshifts $z>1$.
Sometimes, a change of the redshift variable is invoked to address this issue \cite{Cattoen_2007_convergence, Aviles_2012_cosmography_y_variable}. 

The convergence issue in such a kinematic approach is less problematic if Pad\'e rational approximants \cite{Pourojaghi2022, Petreca_2024, Wei_2014_cosmography, Gruber_Luongo_2014_cosmography, Capozziello_Ruchika_Anjan_2019_cosmography, Aviles_2014_cosmography, Mehrabi_2018_cosmography, Rezaei_2017_cosmography, Zhou_2016_cosmography, Liu_2021_cosmography, dutta2020beyond, Capozziello_2019_cosmography, Capozziello_2018_cosmography, Benetti_2019_cosmography}  are used instead of simple power series. The Pad\'e approximant is obtained by expanding a function as a ratio of two power series of order $m$ and $n$ \cite{Pade_1892}. The radius of convergence of such an expansion is usually larger than that of the simple Taylor series expansion \cite{Gruber_Luongo_2014_cosmography}.

In the standard cosmographic approach, the expansion coefficients for Luminosity distance or the Hubble parameter are fitted with data and then subsequently used to reconstruct an effective equation of state $w(z)$ for DE \cite{Adachi_kasai_2012_cosmography, Wei_2014_cosmography, Aviles_2014_cosmography, Gruber_Luongo_2014_cosmography, Zaninetti_2016_cosmography, Zhou_2016_cosmography, Capozziello_Ruchika_Anjan_2019_cosmography, Capozziello_2020_cosmography, dutta2020beyond}.
Using a Pad\'e $(m,n)$ approximation for the Luminosity distance, the expansion coefficients are expressed in terms of kinematic quantities like the present values of the deceleration $q_0$, jerk $j_0$, snap $s_0$ and lerk $l_0$ parameters defined as
\[ 
 q_0 \equiv \frac{-1}{aH^2} \frac{d^2a}{dt^2} \Big|_{a=1},~  j_0 \equiv  \frac{1} {aH^3} \frac{d^3a}{dt^3}\Big|_{a=1}, ~ s_0 \equiv \frac{1}{aH^4} \frac{d^4 a}{dt^4}\Big|_{a=1},~l_0 \equiv \frac{1}{aH^5} \frac{d^5 a}{dt^5}\Big|_{a=1} . \] 
 The constraints on these parameters are then used for the reconstruction of $w(z)$ by either adopting $H_0$, and  density parameters $\Omega_{i0}$  for the known sector (non-dark energy) of the energy budget from other observations or by relating them to these kinematic parameters for a $\Lambda$CDM model. Such an association of the Pad\'e  parameters with parameters of $\Lambda$CDM model gets complicated for higher orders $(m,n)$, or when the Pad\'e approximation is non-trivial involving series expansion in terms of  powers of ${\rm log}(1+z)$ \cite {Lusso_2019} or $\sqrt{1+z}$ \cite{Saini_2000} instead of $z$.

Observations of the background cosmological evolution falls under two broad categories - Measurement of the Hubble expansion rate and measurement of cosmological distances. While they can be independently measured  using cosmic chronometers \cite{Jimenez_2002_CC}, or BAO imprint on the clustering of dark matter tracers \cite{Hu-eisen} or Supernova observations \cite{Scolnic_2022}, they are related to each other. 
Noting that Hubble parameter is related to the derivative of a distance, we are motivated to study cosmological evolution in the phase space.

In this paper,  we formulate the expansion history for $0 \leq z \leq \infty$ in the phase space of the dynamical variables $D_A$ and $d D_A/dz$, where $D_A$ is the angular diameter distance. This is an equivalent formulation to the standard practice of studying $H(z)$ and $D_A(z)$ separately evolving in $z$.
We formulate  a semi-cosmographic method to reconstruct several diagnostics of background cosmological evolution. Starting from a certain pure cosmographic expansion for a measurable quantity say the Luminosity distance, we develop a general way to incorporate non-dark energy model parameters and constrain them simultaneously with the cosmographic parameters using the same data sets.  
The reconstructed evolution history in phase space is  then compared with predictions from some known theoretical  models.

The paper is organized as follows: in Section II we describe the cosmological evolution in phase space and formulate the semi-cosmographic method to reconstruct the background cosmology. In Section III we describe the different observational data sources used to constrain the parameter space. In Section IV we discuss the results and we close with some critical outlook on our work in the concluding Section V.

\section{Formalism} 
\subsection{The Phase-space description}
A comoving length-scale $s$ is expressed as a transverse angular scale $\theta_s = s[(1+z) D_A(z)]^{-1}$ and a radial redshift interval $\Delta z_s = sH(z)/c$, where $D_A(z)$ and $H(z)$ are the angular diameter distance and Hubble parameter respectively. 
By measuring $\theta_s$ and $\Delta z_s$,  both $D_A(z)$ and  $H(z)$ can be determined independently. In Baryon Acoustic Oscillation (BAO) studies, this is achieved by using a standard ruler - the sound horizon $r_d$ at the drag epoch which appears as  the period of oscillation in the transverse and radial clustering of tracers.
The rescaling of distances in the radial and transverse directions also manifests as a source of  anisotropy in  the  redshift space clustering of dark matter tracers through the Alcock-Paczyński (AP) effect \cite{AP1979}. Measurement of this redshift space anisotropy also allows for independent measurement of $D_A(z)$ and $H(z)$.
Instead of working with $(D_A(z), H(z))$, we  shall equivalently consider the phase space  of dynamical variables
\be x (z) = \frac{H_0}{c}D_A(z)~~~~\&~~~p(z) = \frac{H_0}{c} \frac{d D_A}{dz}. \ee
While $D_A(z)$ and $H(z)$ can be independently measured, they are related to each other in a spatially flat cosmology 
through 
\be D_A (z) = \frac{c}{1+z} \int_0^z \frac{dz'}{H(z')}.
\ee
In terms of the dimensionless phase-space variables $(x,p)$,
this gives us a  consistency relationship 
\be x + (1 + z) p  = E(z)^{-1}  
\label{eq:consistency}
\ee
where $E(z) = H(z)/H_0$. 
This implies that for a given redshift $z$, the point $(x(z), p(z))$ must lie on a straight line in the $(x,p)$ phase space given by the above equation. This straight line for the redshift $z$ has a cosmology-independent slope of $-(1+z)^{-1}$ and an intercept on the $p$-axis given by $[(1+z)E(z)]^{-1}$. This relationship acts like a constraint on the phase space. 
The actual phase trajectory is obtained by integrating the dynamical system of the form 
\begin{eqnarray}
& x '& = p , ~~~~~~~
{p}' = F(x, p, z)
\end{eqnarray}
where the \textit{prime} denotes derivatives with respect to the redshift $z$ and the source $F(x,p,z)$ is given by 
\begin{equation} F(x,p,z)  = -\frac{1}{1+z} \left( 2p + \frac{{E}'}{ E^2}  \right). \end{equation}
\begin{figure}[htbp]
\begin{center}
\includegraphics[height=7.5cm, width=7.5cm]{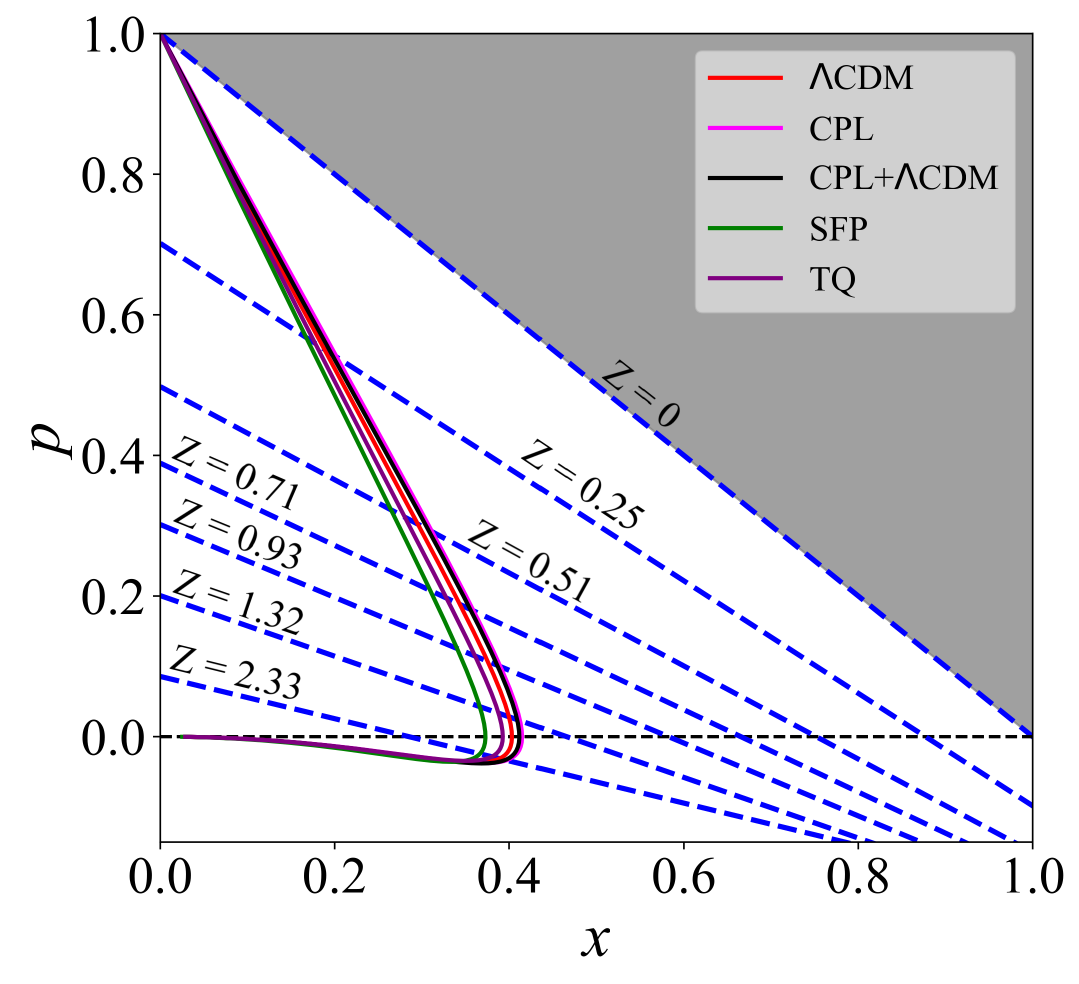}
\end{center}
\captionsetup{font=small} 
\caption{The cosmological evolution in the $(x,p)$ phase space for different models. The dotted straight lines correspond to the consistency condition Eq.( \ref{eq:consistency}) for different redshifts (assuming a $\Lambda$CDM Universe). The intersection of the phase trajectories with these lines gives the value of $(x,p)$ at any particular redshift $z$. }
\label{fig:phasespace}
\end{figure}Thus, to integrate this  dynamical system and thereby determine the evolution history we need the function $E'(z)$.

We will assume that, for a multi-component Universe it is possible to express
\be E(z)^2 = \Omega_{m0} (1+z)^3 + \Omega_{r0}(1 +z)^4 +  \Omega_{\phi} f(z)
\label{eq:split} \ee
where, the dark matter and radiation  components are treated independently  from the unknown DE sector. The function $f(z)$ which imprints the dynamics of DE is usually obtained using the equation of state EoS parameter  $w(z) = P_{DE}/ \rho_{DE}$ for diverse DE models. 

In terms \textbf{of} the equation of state parameter $w(z)$, we have for a spatially flat cosmology 
\begin{eqnarray} 
{E}'= \frac{3 }{2}  \left [  \frac{E ( 1 + w )}{(1+z)} -  \frac{\Omega_{m0} w }{E} (1+ z)^2 + \frac{\Omega_{r0} (1+z)^3}{3E}  - \frac{\Omega_{r0} w (1+z)^3}{E} \right ]  
\end{eqnarray}
Ignoring $\Omega_{r0}$, the source term $F(x,p,z)$ can, thus be written as 
\begin{eqnarray} F(x,p,z) = -\frac{2p}{1+z} - \frac{3}{2} \frac{ \left ( 1 + w \right ) (x + p + pz)}{(1+z)^2}   + \frac{3}{2} \Omega_{m0} w( 1 + z) \left (  { x+ p + pz} \right)  ^3 
\label{eq:source}\end{eqnarray} 
Thus, for a  given DE model $w(z)$ and a set of cosmological parameters, the solution to the problem of finding the evolution history effectively reduces to solving a three dimensional autonomous system of non-linear differential equations
\begin{eqnarray}
& x'& = p  \nonumber \\
&{p}'& =F(x,p,z)  \nonumber \\
&z'& = 1
\label{eq:autonomous}
\end{eqnarray}
with initial conditions $(x_0, p_0, z_0) = ( 0, 1, 0) $.
The solution to this system of equations directly gives us $(x(z), p(z))$ which can, then be used to find $D_A(z)$ and $H(z)$ using the consistency relation in Eq.(\ref{eq:consistency}). 
Figure \ref{fig:phasespace} shows the cosmic expansion history in the $(x,p)$ phase space.
The lines of consistency in Eq.(\ref{eq:consistency}) are shown in the figure for some redshifts. 
The present epoch $(z=0)$ corresponds to the line that passes through $(0,1)$ and $(1,0)$ and all epochs  with $z>0$  lie below it. 
The shaded gray region above this line corresponds to $z<0$.
The intersection of the straight line corresponding to any redshift $z$ with the solution $(x(z),p(z))$ of the dynamical system in Eq.(\ref{eq:autonomous}) gives the  value of $(x,p)$ at that redshift.
Regardless of the cosmological model, all phase trajectories must start at $(0,1)$ at $z=0$ and approach the origin $(0,0)$ as $z \rightarrow \infty$ which corresponds to the big bang. There is a redshift at which the phase trajectory intersects the $p=0$ line, which corresponds to the maximum angular diameter distance.
The difference between different cosmological models is maximal around this redshift in the $(x,p)$ plane.
The phase trajectory can be obtained by adopting some model $w(z)$.
In figure \ref{fig:phasespace}, we have shown the cosmological evolution for the following models in phase space.
\begin{itemize}
\item The $\Lambda$CDM model ($w =-1$): For this widely popular model, we have adopted the cosmological parameters {\textbf from Planck \cite{aghanim2020planck}}.
\item 
The 
CPL model: In this model \cite{CHEVALLIER_2001}  $w_{\phi}$ is given by 
\be w_{\phi}(z) = w_0 + w_a \frac{z}{1+z} \ee and provides a phenomenological  parametrization
to describe several features of DE using two parameters $(w_0, w_a)$. This model has  
been extensively used  as the standard two parameter  characterization of dynamical DE \cite{albrecht2006report}. Further,  a wide class of quintessence
scalar field models can  also effectively  be mapped into the CPL parametrization \cite{PhysRevD.93.103503}. As $z\rightarrow 0$, $w \rightarrow w_0$ and for the early Universe as $z \rightarrow \infty $,  $w \rightarrow w_0 + w_a$. This also means that the parameter $w_a$ can not take a very wide range of values in this model.

\item The CPL-$\Lambda$CDM model: This is  a cosmological model involving a negative cosmological constant (AdS vacua in the DE sector) along with a quintessence field ($\rho_{DE} = \Lambda + \rho_{\phi}$)
The quintessence field is given a CPL parametrization ($w_0, w_a$). The parameters of this model are adopted from \cite{dash2024post}. 

\item Scale Factor Parametrization (SFP): In this model the scale factor is parametrized using two parameters $A$ and $B$ in a way that 
\[ H^2(z) = H_0^2 \left [ A( 1+z)^{2/B} + ( 1- A) \right ]. \]
In this model all the observables related to background
evolution are constructed from the scale factor $a(t)$ only \cite{scalefactorparam}.

\item{Thawing Quintessence (TQ):}
Thawing models are characterized by flat potentials and the field begins with $w \sim -1$  and increases only slightly to the present epoch. We adopt the equation of state from 
\cite{PhysRevD.77.083515}. 
\end{itemize}
This is only a small sample of DE models,  and by no means exhaust the  range of possible dynamical equations of state $w(z)$.
In this work, instead of assuming any specific model $w(z)$, we adopt a kinematic approach.  We shall discuss this in the next section.

\subsection{The semi-cosmographic reconstruction}
We adopt two kinematic cosmographic descriptions of background evolution where the  Luminosity distance may be   expressed as a Pad\'e rational fraction expansion.
Firstly, we consider the standard Pad\'e approximants as ratios of polynomials in the redshift $z$ \cite{Liu_2021_cosmography, Aviles_2014_cosmography, Capozziello_2020_cosmography, Hu_2022_cosmography, Gruber_Luongo_2014_cosmography}  and assume that
\be {\rm \bf Case~ I} : ~~~~~~ D_L^{\cal {P}} (z) = \frac{c}{H_0} {\mathcal{P}}_{(m,n)}(z)
\label{eq:pade_P}
\ee
where $\mathcal{P}_{(m,n)}$ is the Pad\'e approximant $(m,n)$ given by 
\be 
{\cal {P}}_{(m,n)}(z) = 
\frac{ a_1 z + a_2 z^2 + \dots + a_m z^m}{1 + b_1 z + b_2 z^2 + \dots + b_n z^n}. \ee
The expansion coefficients can be related to kinematic quantities like $H_0$, $q_0$, $j_0$, $s_0$, $l_0$, ... by comparing this Pad\'e expansion with the Taylor expansion \cite{Aviles_2012_cosmography_y_variable, Petreca_2024, Capozziello_lazkoz_2011_cosmography, Liu_2021_cosmography, Hu_2022_cosmography, Gruber_Luongo_2014_cosmography, Capozziello_Ruchika_Anjan_2019_cosmography, Aviles_2014_cosmography, Weinberg_1972_cosmography, Capozziello_2020_cosmography}. We don't make any such connections \textit{ab initio}, and treat the Pad\'e expansion coefficients $(a_m, b_n)$ themselves as the parameters of interest. 

Moreover, we note that connections of the Pad\'e parameters with the kinematic quantities become more complicated in the second description that we adopt.
In this case, we assume that the Pad\'e expansion is in terms of the variable $\xi = \sqrt{1+z}$ \cite{Saini_2000}.
\be {\rm \bf  Case~ II} : ~~~~~~D_L^{\cal{R}} (z) = \frac{c}{H_0} {\cal {R}} (\sqrt{1+z})
\label{eq:pade_R}
\ee
where the function ${\cal{R}} (\xi)$ is given by 
\be 
{\cal{R}} (\xi) = 2 \left [ \frac{  \xi^4 - \alpha  \xi^3 -( 1- \alpha )\xi^2 }{ \beta \xi^2 + \gamma  \xi + 2 - \alpha  - \beta - \gamma }   \right ]. 
\ee
This choice of the Pad\'e approximant is motivated by the fact that $D_L$ and $H$ obtained from it have the desired  asymptotic behaviour \cite{Saini_2000}.

For each of these cosmographic  descriptions ({\rm \bf  Case~I} and {\rm \bf Case~II}), the corresponding approximants for angular diameter distance $D_A^{{\cal P}/{\cal R}}(z)$ and Hubble parameter $H^{{\cal P}/{\cal R}}(z)$ for a spatially flat cosmology are given by 
\be  D_A^{{\cal P}/{\cal R}} (z)   = \frac{ D_L^{{\cal P}/{\cal R}}}{(1+z)^2} ~~\&~~ H^{{\cal P}/{\cal R}} (z)= c\left [ \frac{d}{dz} \frac{D_L^{{\cal P}/{\cal R}}}{(1+z)}\right ]^{-1}. \label{eq:pade1}\ee
In the standard cosmographic approach, $D_L^{{\cal P}/{\cal R}}(z)$,  $D_A^{{\cal P}/{\cal R}}(z)$ and $H^{{\cal P}/{\cal R}}(z)$ are fitted with observational data to obtain constraints on kinematic parameters $( a_1, a_2 \dots, a_m,  b_1, b_2 \dots, b_n, H_0) $ or $( \alpha, \beta, \gamma, H_0)$ depending on which case one adopts. This does not allow for easy direct constraints on $\Omega_{m0}$ or any other density parameters, unless the kinematic quantities are expressed in terms of these parameters using the Taylor-Pad\'e connection. Instead of trying to connect the Taylor expansion with the Pad\'e expansion, we propose a semi-cosmographic approach to constrain the expansion history by defining 
\be w^{[\Omega_m,{\cal P}/{\cal R}]}\left(z\right) = \frac{\frac{2}{3}\left(1+z\right)\frac{d}{dz} {\rm{ln}} H^{{\cal P}/{\cal R}}(z) - 1}{1 - \left ( \frac{H_{0}}{H^{{\cal P}/{\cal R}}\left (z \right )} \right ) ^2\Omega_{m0}\left(1+z\right)^3}.
\label{eq:eos}
\ee
This semi-cosmographic equation of state parameter $ w^{[\Omega_m,{\cal P}/{\cal R}]}\left(z\right)$, now additionally depends on $\Omega_{m0}$ along with the kinematic parameters $(H_0, a_1, a_2 \dots, a_m, b_1, b_2 \dots, b_n)$ of ${\cal P}_{(m,n)}(z)$ or
 $(H_0, \alpha, \beta, \gamma)$ of ${\cal R}(z)$. It seams together kinematic information in  ${\cal P} (z)$ or ${\cal R}(z)$  with the dynamical information imprinted in $E(z)$.
Using this semi-cosmographic equation of state $w^{[\Omega_m,{\cal P}/{\cal R}]}$ in Eq.(\ref{eq:source}),  the autonomous system in Eq.(\ref{eq:autonomous}) can be solved numerically for $x(z)$ and $p(z)$. 
This gives us a new set of quantities $(D_L(z), D_A(z), H(z))$ for each of the  scenarios $\bf I $ and $\bf II$:
\begin{eqnarray} 
D_L(z)= \frac{c(1+z)^2 x}{H_0}, ~
D_A(z) = \frac{cx}{H_0},~
H(z)= \frac{H_0}{x + p + pz}.
\end{eqnarray}
Thus, for each of the starting cosmographic scenarios, we have two sets of expressions for the distances and Hubble expansion rate: 
the original cosmographic $(D_L^{{\cal P}/{\cal R}}, D_A^{{\cal P}/{\cal R}}, H^{{\cal P}/{\cal R}})$ and $(D_L, D_A, H)$.
The first set has no pre-assumptions about cosmological dynamics, whereas the second set is based on the specific form of $E(z)$ in Eq.(\ref{eq:split}) and the equation of state $w^{[\Omega_m,{\cal P}/{\cal R}]}$ in Eq.(\ref{eq:eos}). For internal consistency, observational data must constrain the parameter space simultaneously for both sets of functions for the same physical quantities. Given the distinct nature of the two forms of functions, the posterior distribution of the parameters for joint estimation using both together tends to give a bimodal distribution for $\Omega_{m0}$. To avoid this issue, we use the posterior distribution on the parameters $(H_0, a_1, a_2,  \dots, b_1, b_2, \dots)$ or $(H_0, \alpha, \beta, \gamma)$ obtained by directly fitting data with $(D_L^{{\cal P}/{\cal R}}, D_A^{{\cal P}/{\cal R}}, H^{{\cal P}/{\cal R}})$ as priors for fitting the same data with $(D_L, D_A, H)$. For this second fitting based on the solution of the autonomous system with $w^{[\Omega_m,{\cal P}/{\cal R}]}$, we additionally assume flat priors for $\Omega_{m0}$. The fit using 
 $(D_L^{{\cal P}/{\cal R}}, D_A^{{\cal P}/{\cal R}}, H^{{\cal P}/{\cal R}})$ limits the vast parameter space of the kinematic parameters, while the subsequent fit using $(D_L, D_A, H)$, shifts the parameters for consistency with dynamical information now incorporated in the modeling. 

Using cosmological data on distances and Hubble parameter and adopting this two step fitting process, the phase-space orbit can be reconstructed.
Apart from that, the best fit values of the parameters and their respective errors are used to reconstruct some important diagnostic probes of background cosmology. We apply our fitting method to reconstruct the following quantities which are related to each other.

\begin{itemize}
\item { {\bf The dark energy EoS $w(z)$:} }
This is the most commonly used quantifier of DE dynamics. For accelerated expansion we require that DE violate the strong energy condition with $w(z) < -1/3$. If the acceleration is driven by the cosmological constant then $w = -1$ and any departure from this implies that DE is dynamic.
For scalar field DE one may have the freezing models, where $w(z)$ does not evolve significantly and remains close to $-1$ throughout the cosmic expansion. In case of thawing models $w(z)$ starts close to $-1$ and evolves toward less negative values as the universe expands.

\item { \bf The ${\cal{O}}m$ diagnostic:} The ${\cal{O}}m$ diagnostic proposed in \cite{Sahni_2008_Om}, is an useful quantifier of evolving DE. This is defined as
\be
{\mathcal{O}}  m (z)= \frac{  E(z)^2  - 1}{\left(1+z\right)^3 -1}.
\ee
Except for $z=0$, where it diverges, this quantity measures any departure from the $\Lambda$CDM  model since its value is a constant $\Omega_{m0}$ for the  $\Lambda$CDM  model.
Further,  ${\mathcal{O}}  m (z) > \Omega_{m0}$ in Quintessence and ${\mathcal{O}}  m (z) < \Omega_{m0}$ in Phantom DE models.

\item{\bf Evolution of DE $f(z)$:} We define DE evolution using 
\begin{equation}
    f(z) = \frac{E(z)^2 - \Omega_{m0}\left(1+z\right)^3}{1 - \Omega_{m0}}.
\end{equation}
Like ${\mathcal{O}}  m (z)$ and $w(z)$, this quantity imprints the evolution of the DE density. For the $\Lambda$CDM model, $f(z) =1$.
Thus, any departure from unity at low redshifts indicates dynamical  DE.
\item{ \bf The AP distortion parameter $F(z)$:}
The Alcock–Paczynski effect \cite{AP1979} is used to constrain  cosmological models by comparing the observed tangential and radial size of objects which
are otherwise assumed to be isotropic. If $\Delta z$
and $\Delta \theta$ are the radial and tangential extents of the object then the quantity of interest is $F(z) = \Delta z/ \Delta \theta$. This can be written as
\begin{equation} 
    F(z) = \frac{(1+z)D_A(z)H(z)}{c}.
\end{equation}
The AP test aims to measure the departure of this quantity from its value in a fiducial cosmology.

\item{ \bf The BAO distance measure $D_V(z)$:}
Galaxy surveys imprint  both the transverse and the radial BAO peaks. It is however difficult to probe large radial distances leading  to small survey depths. Further, large shot noise degrades the SNR making it very difficult to independently measure $D_A(z)$ and $H(z)$. Typically, the combination $D_V(z)$ is measured instead in galaxy
redshift surveys \cite{Eisenstein_2005, Percival_2007} given by 
\begin{equation}
    D_{V}(z) = \left[ (1+z)^2 D_{A}(z)^{2} \frac{cz}{H(z)} \right]^{\frac{1}{3}}.
\end{equation}
This quantity is often used in BAO analysis when high SNR anisotropic data is not available.
\end{itemize}
\section{Observational Aspects and Data}
In this section, we discuss the cosmological data used from various cosmological probes for our analysis.
We have considered three main data sources. Since our initial Pad\'e expansion is for the Luminosity distance, we consider distance measurements using SNIa apparent magnitude. We also consider BAO data which gives the $(D_A(z), H(z))$ information for our phase space analysis and cosmic chronometer (CC) data for $H(z)$ measurements.

\subsection {BAO Data}
 
 We use the BAO data on $\tilde{D}_{M}$  and $ \tilde{D}_H$
    defined as 
    \be \tilde{D}_{M} = \frac{c}{r_d} \int_0^z \frac{dz'}{H(z')}  ~~~\&~~~~  \tilde{D}_{H} = \frac{c}{H r_d} \ee
    where $r_d$  is the sound horizon at the drag epoch. In our analysis we have adopted $r_d = 146.995 \pm 0.264~ {\rm Mpc}$, from CMBR constraints \cite{CMB-distance-prior, Dinda_2025}.
    The tracers included in the DESI BAO data are luminous red galaxy (LRG), emission line
galaxies (ELG) and the Lyman-$\alpha$ forest (Ly-$\alpha$QSO) in
a redshift range $0.1 \leq  z \leq 4.2$.  
   We adopt the anisotropic BAO  data from DESI DR1 \cite{DESI_2025_DR1} for LRG, ELG and Ly-$\alpha$ tracers at $5$ redshifts $z_{eff} =0.51, 0.706, 0.93, 1.32$ and $2.33$ respectively.  The mean, variance and correlation information is adopted from \cite{Adame_DESI_Data_2025}. 
    For two other redshifts $z_{eff} =0.295$ and $1.491$  we have taken the isotropic BAO data on $D_V$ \cite{Adame_DESI_Data_2025}.    
    All the given covariance matrices are suitably transformed using the Jacobian for the corresponding transformations to obtain the covariance matrix for the  relevant quantities $(x,p)$, whenever required. 

\subsection{SNIa Data}
 The measurement of the Luminosity distance using Supernova Type Ia (SNIa) has been a crucial cosmological probe and amongst the earliest to indicate cosmic acceleration \cite{perlmutter1998discovery}.
The Pantheon+ sample consists of apparent magnitude data for  1701  SNIa light curves  in the redshift range
$0.00122 \leq z \leq 2.26137$ \cite{Scolnic_2022, Brout_2022_pantheon}. To avoid the issue of strong peculiar velocity dependence at low redshifts \cite{Brout_2022_pantheon} we have not considered $111$ light curves in the range $z < 0.01$. We have adopted the data and its full statistical and systematic covariance from \url{https://github.com/PantheonPlusSH0ES/DataRelease}.

 \subsection{Cosmic Chronometer (CC) Data}

The cosmic chronometers (CC) method is a model independent method to measure $H(z)$ \cite{Jimenez_2002_CC, Stern_2010_CC, Ratsimbazafy_2017_diff_age, Zhang_2014_SDSS_CC} by using the relation 
\be 
H(z) = - \frac{1}{1+z} \frac{dz}{dt}. \ee
While, redshifts can be measured with high precision uisng spectroscopic techniques, the main 
difficulty in this method is to accurately determine $dt$, the differential age evolution. This requires cosmic chronometers. Passive stellar
populations and passive early type galaxies are some good CC candidates. We use the data on $32$ CCs \cite{CC-Moresco-2020} in the redshift range $0.07 \leq z \leq 1.965$ and the full covariance matrix from \url{https://github.com/Ahmadmehrabi/Cosmic_chronometer_data}.

\subsection{BAO imprint on the 21-cm Intensity Mapping}

Traditional Baryon Acoustic Oscillation (BAO) surveys, such as DESI and BOSS, rely on the distribution of galaxies and quasars, which are typically constrained to redshifts  $(z \leq 3)$. In contrast, 21 cm Intensity Mapping (IM) probes BAO deep into the reionization era $( z > 6 )$, extending BAO studies further into cosmic history \cite{Liu:2019awk}. IM offers significant advantages over galaxy surveys due to its ability to cover larger volumes at higher redshifts, thereby improving the precision of the results \cite{bharad04, Bull_2015}. Several radio telescopes such as HIRAX \cite{Crichton:2021hlc}, CHIME \cite{CHIME:2022dwe}, and SKA \cite{Santos:2015gra, Xu:2020uws} are aiming to detect BAO signal using 21cm IM in near future. The potentially large survey volumes for 21-cm intensity mapping experiments makes it possible to measure radial and transverse BAO features with high SNR.
In the absence of actual data, we model the observed data using error projections from a
SKA1-Mid like radio interferometer.
 
In the post-reionization universe $( z \leq 6 )$, DLAs are the dominant reservoirs of HI, containing $\approx 80\%$ of the neutral hydrogen at  $z < 4$  \citep{proch05} with HI column density greater than $ 2 \times 10^{20}$atoms/$\rm cm^2$ \citep{xhibar, xhibar1, xhibar2}. 
The post EoR power spectrum $P_{21}$ of the 21-cm excess brightness temperature field can be modeled in the linear regime as \citep{Furlanetto_2006, Bull_2015, bharad04, param3}
\begin{equation}
P_{21}(k, z, \mu)  = C^2_T(z) ~ {( 1  + \beta_T \mu^2)}^2  ~ P_m(k, z)
\end{equation}
where $\mu = {\bf{k}} \cdot {\bf{\hat{n}}}$ and $\beta_T = f_g(z)/b_T$, where $f_g(z)$ is the logarithmic  growth rate of matter fluctuations,  $b_T$ being the HI bias and $P_m$ is the dark matter power spectrum \cite{Hu-eisen}. The redshift space distortion (RSD) factor $1+\beta_T\mu^2$ arises due to the  peculiar velocity of the HI clouds  \citep{poreion2, bharad04, kaiser1987clustering}. The overall amplitude ${\mathcal C}_T$ is the average HI brightness temperature, given by
\begin{equation}
{\mathcal{C}}_{T} = 4.0 \, {\rm {mK}} \,
b_{T} \, {\bar{x}_{\rm HI}}(1 + z)^2\left ( \frac{\Omega_{b0}
  h^2}{0.02} \right ) \left ( \frac{0.7}{h} \right) \left(
\frac{H_0}{H(z)} \right).
\end{equation}
The mean HI fraction $\bar{x}_{\rm HI}$ and bias $b_T$ that completely model post-EoR 21cm power spectrum are largely uncertain. However, in the  post-EoR epoch $\bar{x}_{HI}$ does not evolve much \cite{xhibar, xhibar2}. Simulation studies show that the bias is scale dependent on small scales below  the Jean's length \citep{fang}. However, on large scales the bias is expected to be scale-independent \cite{Guha_Sarkar_2012, Sarkar_2016, Mar_n_2010}. In our analysis we kept the fiducial value of $\bar{x}_{HI} = 2.45 \times 10^{-2}$ \cite{xhibar2} and consider the fitting of bias from \cite{Sarkar_2016}.
The BAO manifests itself as a series of oscillations in the linear matter power spectrum. The Baryonic feature is seen clearly if  we subtract the cold dark matter contribution from the total power spectrum: $P_b (k) = P(k) - P_c(k)$. The BAO power spectrum can be modeled as
\citep{hu1996small, seo2007improved} \begin{equation}
\label{eq:baops}
P_b (k') = A \frac{\sin x}{x} e^{-(k'\Sigma_s)^{1.4}}e^{-k'^2
  \sum_{nl}^2/2}
\end{equation}
where $A$ is a normalization constant, $\Sigma_s = 1/k_{silk}$ and $\sum_s = 1/k_{nl}$ denotes the inverse scale of \lq
Silk-damping' and \lq non-linearity' respectively. In our analysis we
have used $k_{nl} = (3.07 h^{-1}$Mpc$)^{-1} $and $k_{silk} = (8.38
h^{-1}$Mpc$)^{-1} $ from \cite{seo2007improved} and $x =
\sqrt{k^2(1-\mu^2) s_\perp^2 + k^2\mu^2 s_\parallel^2}$, where  $s_{\perp}$  and  $s_{\parallel}$ are the transverse and radial sound horizon scales, respectively. The changes in $ D_A(z)$  and  $H(z)$  are reflected in  the variations of  $s_{\perp}$  and  $s_{\parallel}$. The fractional errors in these quantities correspond to the uncertainties in  $D_A/s$  and  $sH$, where  $s= r_d$  represents the true physical value of the sound horizon.

To quantify these errors, we define the parameters, $p_1 = \ln (s^{-1}_{\perp})$ and $p_2 = \ln (s_{\parallel})$ and use them in our analysis to derive the Cramer-Rao bounds:
$\sqrt{F^{-1}_{11}} = \delta D_A/D_A$ and $\sqrt{F^{-1}_{22}} = \delta H/H$ respectively, where  $F_{ij}$ represents the Fisher matrix elements and is given by \cite{sarkar2013predictions}
\begin{eqnarray}
    F_{ij} =  \int dk' ~ \int_{0}^{1} d\mu ~ \frac{C^2_T}{\delta P^2_{21}} \left[ 1 + \beta_T \mu^2 \right]^2  \left( \cos x - \frac{\sin x}{x}\right)^2  \nonumber \\ \times f_i(\mu) f_j(\mu) A^2 e^{-2(k'\sum_s)^{1.4}}e^{-k'^2  \sum_{nl}^2}  \nonumber
\end{eqnarray}
where $f_1 = \mu^2 -1$ and $f_2 = \mu^2 $. The term $\delta P_{21}^2$ is the variance of 21cm experiment. We adopt the theoretical expected noise for a radio interferometric experiment from \cite{villaescusa2014modeling, Sarkar_2015, geil2011polarized, param4}.
For a radio interferometric observational frequency $\nu = 1420/(1+z)$MHz or wavelength $\lambda = 0.21(1+z)$m, we have 
\begin{equation}
\delta P_{\rm HI}(k, \mu, z)=\frac{P_{\rm HI}(k,\mu,z) +N_T(k, \mu , z)}{\sqrt{N_c}}
\end{equation}
where 
\begin{equation}
N_T=\frac{\lambda^2 T_{s y s}^2 r^2 d r / d \nu}{A_e t_{\mathbf{k}}}.
\end{equation}
Here $A_e$ is the effective area of the individual antenna dish, $T_{sys}$ is the system temperature, $r$ is the comoving distance to the source and 
\begin{equation}
t_{\mathbf{k}}=T_0 N_{a n t}\left(N_{a n t}-1\right) A_e \rho / 2 \lambda^2  \end{equation}
is the fraction of the total observation time $T_0$ spent on each mode.  We have considered a radio-array with $N_{ant}$ antennae spread out in a plane, such that the total number of visibility pairs $N_{ant} ( N_{ant} -1) /2$ are distributed over different baselines according to a normalized baseline distribution function $\rho(k_{\perp}, \nu)$
\begin{equation}
\rho_b \left({\bf{k_{\perp}}} =\frac{2 \pi {\bf u}}{r}\right) = c \int d^2 {{\bf r}} \rho_{ant} ({\bf r}) \rho_{ant} ({\bf r} - \lambda {\bf{u}} ). \end{equation}
Where $c$ is fixed by normalization of $\rho_b({\bf{u}})$ and $\rho_{ant}$ is the  distribution of antennae. We assume 
$\rho(r) \sim 1/r^2$.

The noise is suppressed by a factor $\sqrt {N_c}$ where $N_c$ is the number of modes in a given survey volume. We have 
\begin{equation}
N_c=2 \pi k^2 \Delta k \Delta \mu r^2(dr / d \nu) B \lambda^2 / A_e(2 \pi)^3 .   
\end{equation}
The noise estimates are based on a futuristic SKA1-Mid like intensity mapping experiment. We consider an interferometer with $250$ dish antennae each of diameter $15$m
For the SKA-Mid frequency band 1 and 2 ($400-950$MHz) the assumed frequency bandwidth is $32$ MHz. We assume $500$ hours of observation per pointing and consider multiple pointings for a full sky observation. We consider the spherically averaged  power spectrum which is binned in logarithmically  spaced bins in $k$, with $dk/k = 1/6$. The minimum wavenumber is set to $k_{\rm min} = 0.005$ h Mpc$^{-1}$ to ensure the validity of Newtonian perturbation theory, while the maximum wavenumber is limited to $k_{\rm max}=0.2$ h Mpc$^{-1}$ to remain within the linear regime. For sufficiently long observations, instrumental noise becomes negligible, and the signal-to-noise ratio (SNR) is dominated by cosmic variance. In this regime, the covariance of the measurement can only be further reduced by a factor of $1/\sqrt{N_{p}}$ where $N_{p}$ is the number of independent pointings. We also note that small $k_{\parallel}$, are plagued by foreground contaminants. This corresponds to the large $\Delta_\nu$ over which the foregrounds are correlated, requiring  us to remove these small $k_{\parallel}$ modes. 

\begin{figure*}[htb]
\begin{center}
\includegraphics[height=9cm, width=9cm]{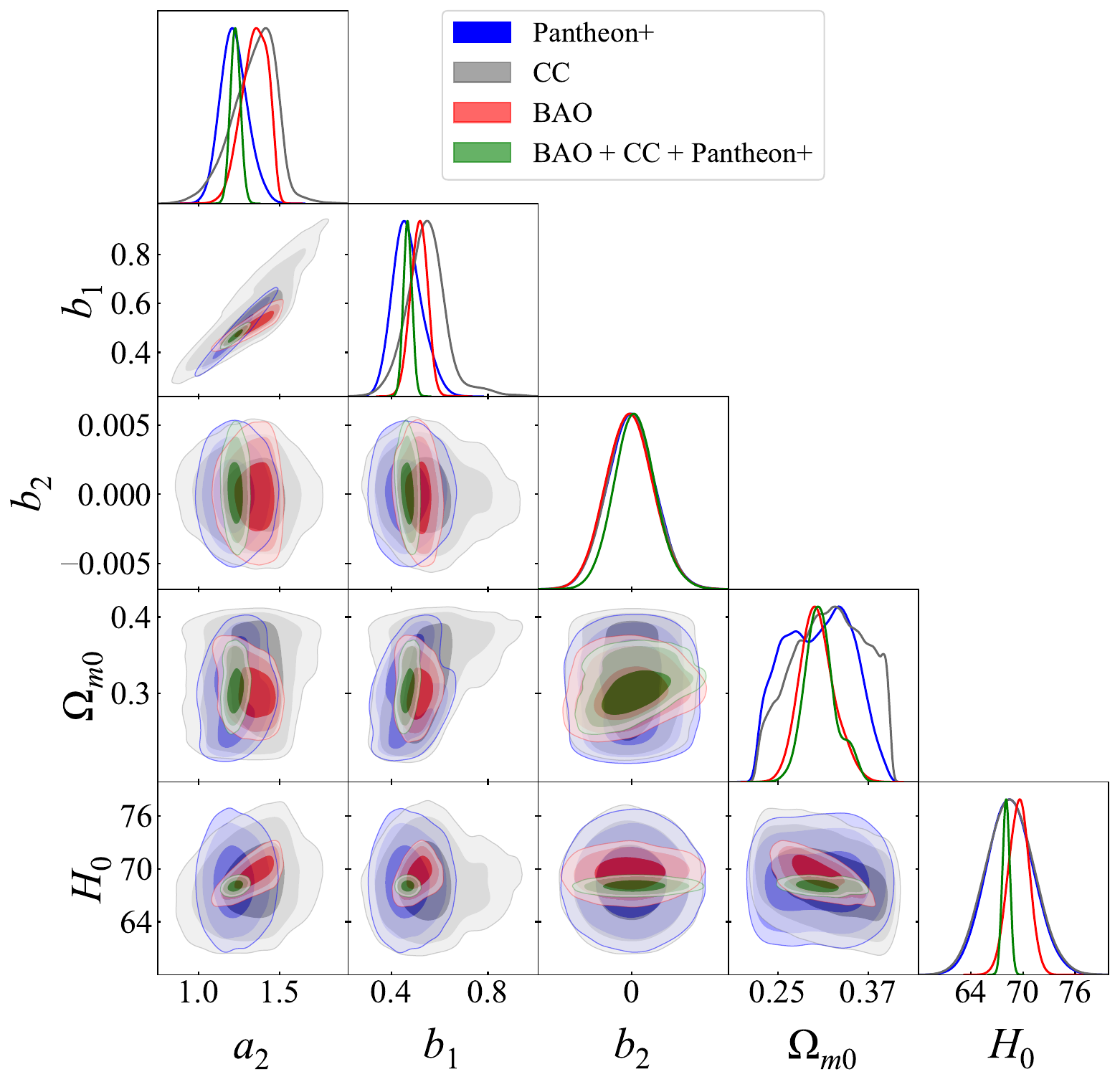}
\includegraphics[height=7.5cm, width=15cm]{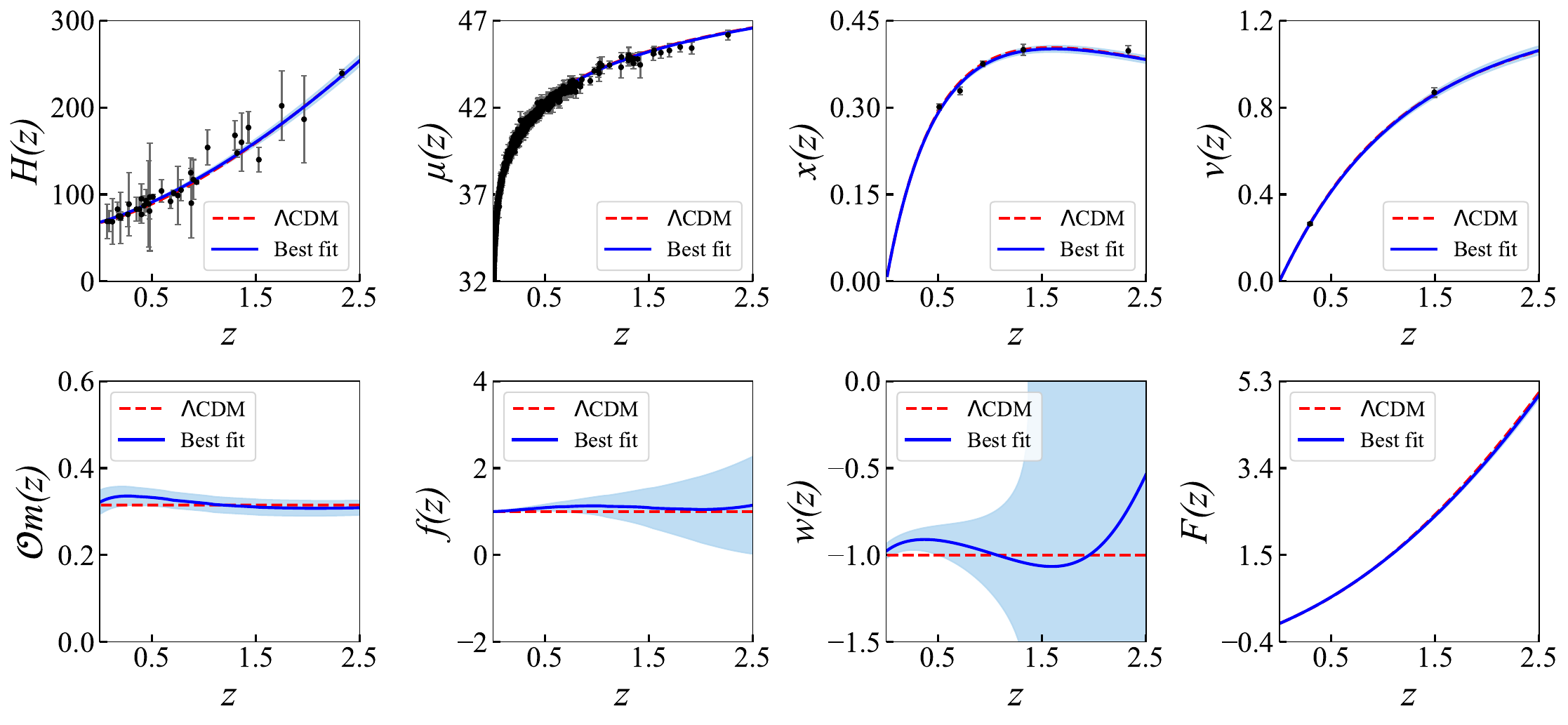}
\captionsetup{font=small} 
\caption{Marginalized posterior distribution of the set of semi-cosmographic parameters ($H_0$, $\Omega_{mo}$, $a_2$, $b_1$, $b_2$) and the corresponding 2D confidence contours obtained from the MCMC analysis starting with $D_L^{\mathcal{P}}$. 
The confidence contours correspond to data being fitted with BAO (DESI DR1), CC,  Pantheon+, and joint (BAO + CC + Pantheon+), respectively.
The panel below  shows the reconstruction of some diagnostics of background cosmology and their $1\sigma$ errors from the joint (BAO + CC + Pantheon+) analysis. }
The corresponding quantities for the $\Lambda$CDM model are  superposed for comparison.

\label{fig:reconstruct-P}
\end{center}
\end{figure*}

\begin{figure*}[htb]
\begin{center}
\includegraphics[height=7cm, width=7cm]{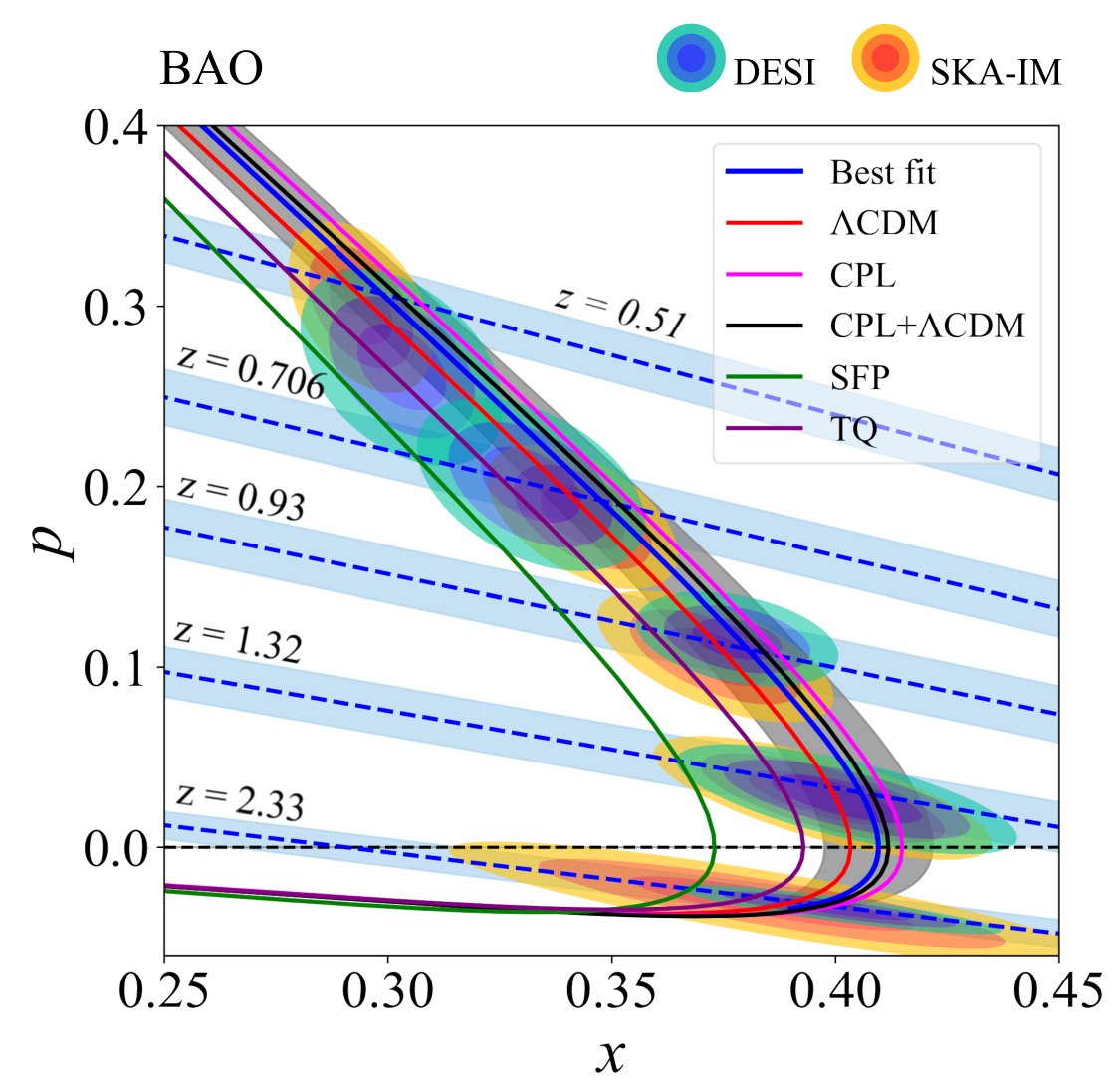}
\includegraphics[height=7cm, width=7cm]{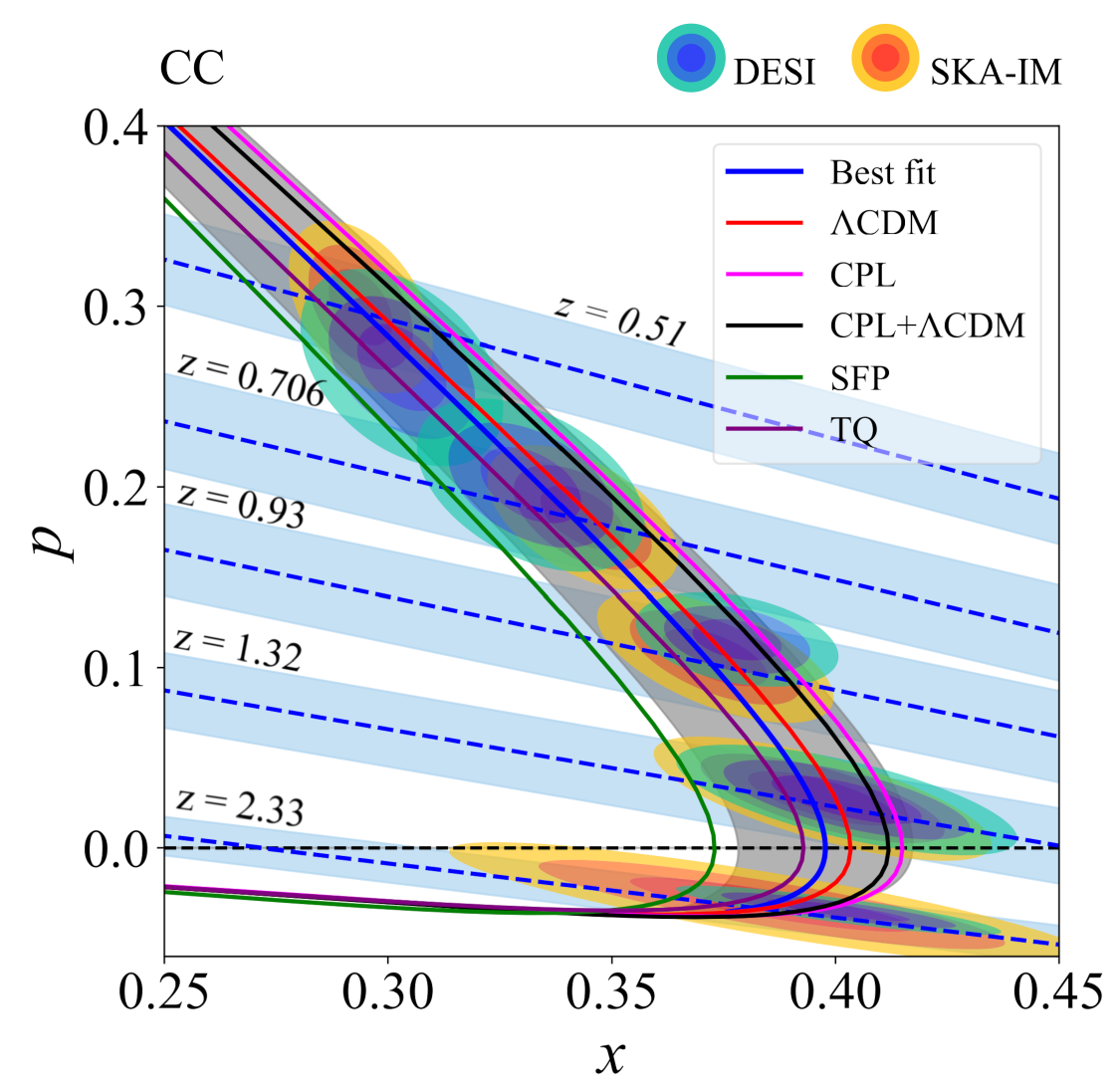}
\includegraphics[height=7cm, width=7cm]{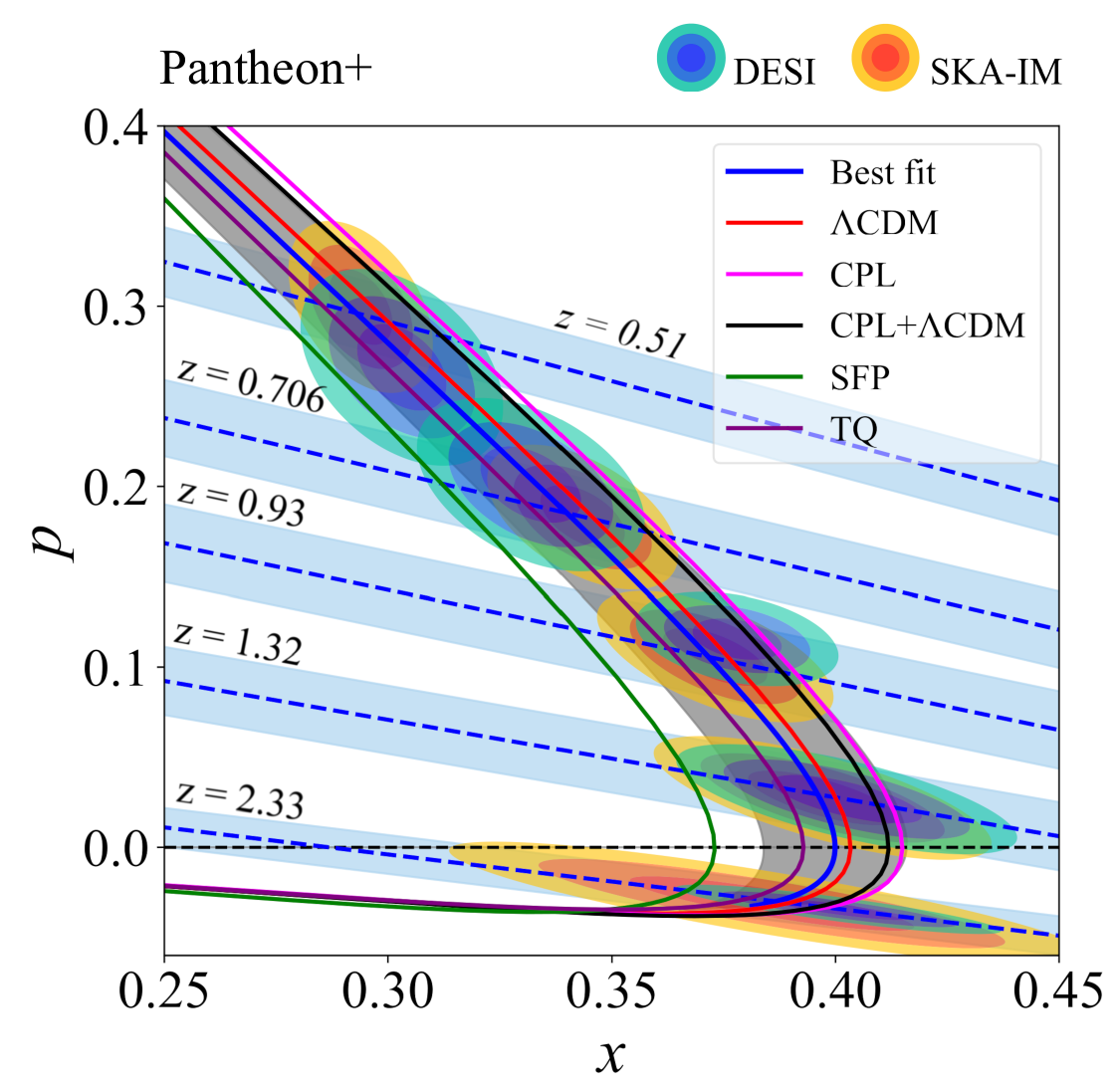}
\includegraphics[height=7cm, width=7cm]{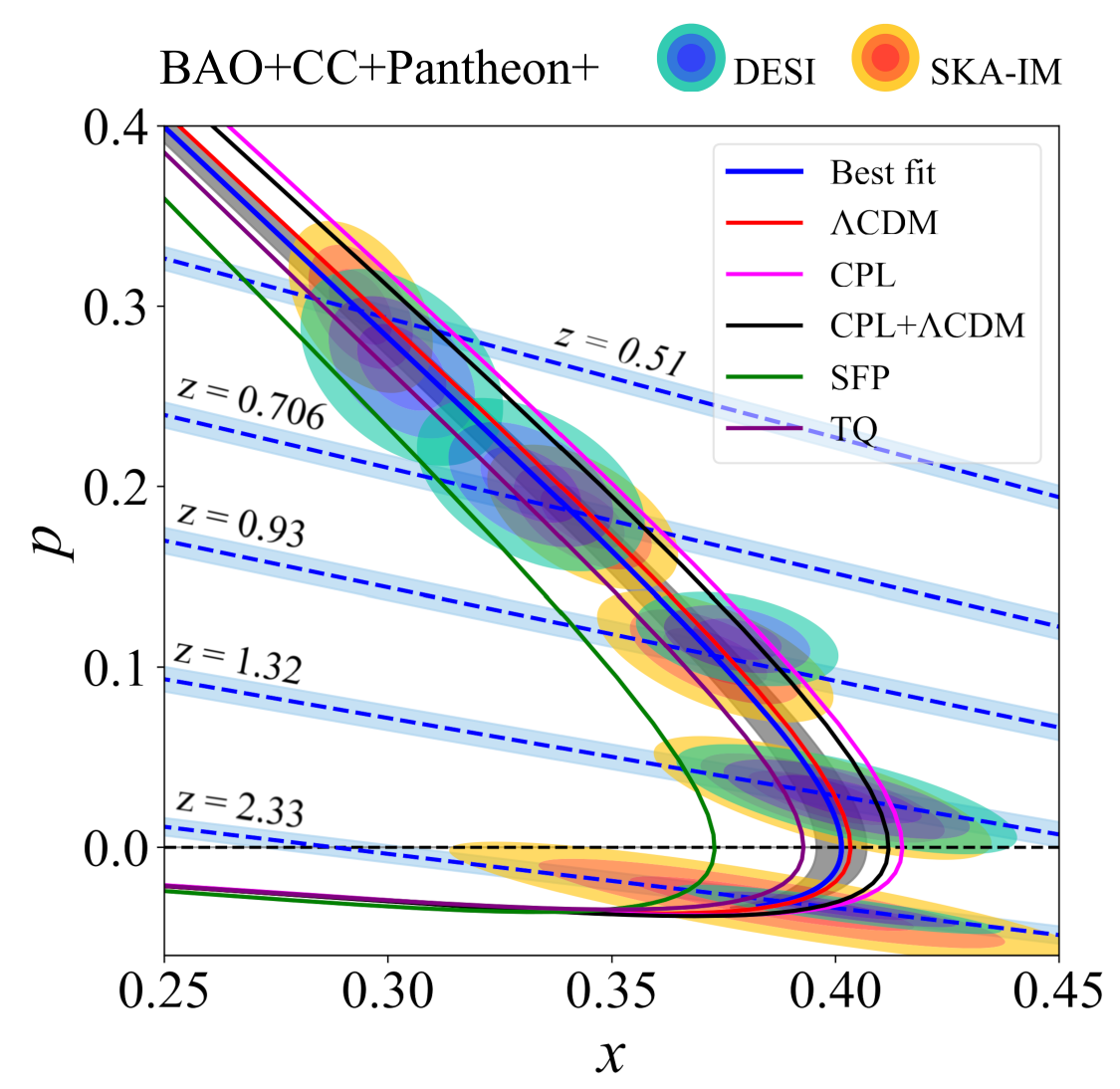}
\end{center}
\captionsetup{font=small} 
\caption{The reconstructed phase space trajectory $(x(z), p(z))$ with $1\sigma$ error for cosmography starting with $D_L^{{\cal P}}(z)$. Several DE models are also shown for comparison. 
The top-left figure corresponds to a reconstruction with BAO data, the top-right figure corresponds to CC data. The lower-left figure corresponds to a reconstruction using Pantheon+ data and the lower-right corresponds to a  joint analysis. 
The actual DESI error contours at 5 redshifts and the projected error contours for a 21-cm intensity mapping experiment at observing frequencies corresponding to the same redshifts are shown. We also show the consistency lines corresponding to the same redshifts with its error (originating from $H(z)$).}
\label{fig:phasespace-rec-P}
\end{figure*}

\begin{figure*}[htb]
\begin{center}
\includegraphics[height=9cm, width=9cm]{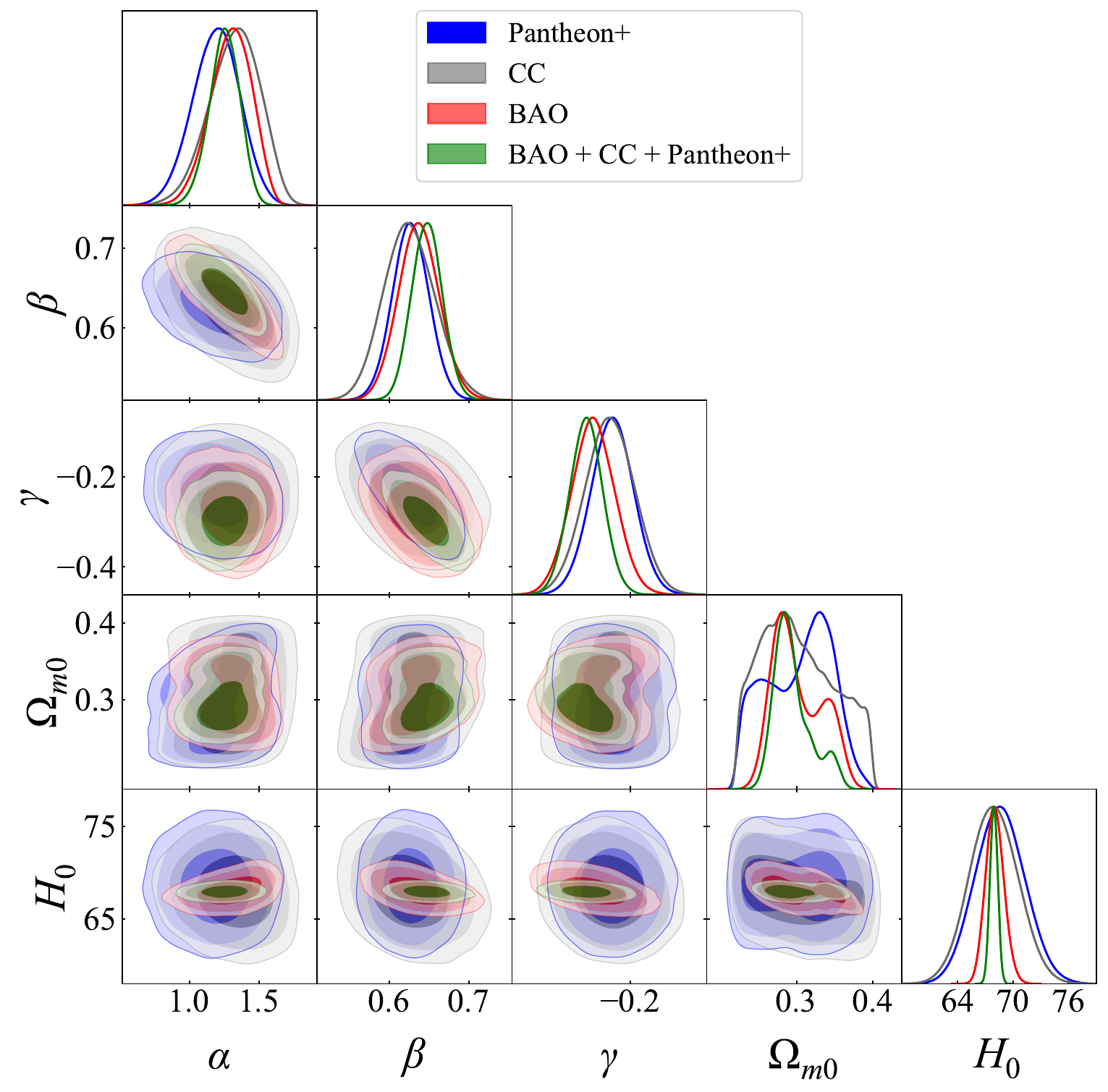}
\includegraphics[height=7.5cm, width=15cm]{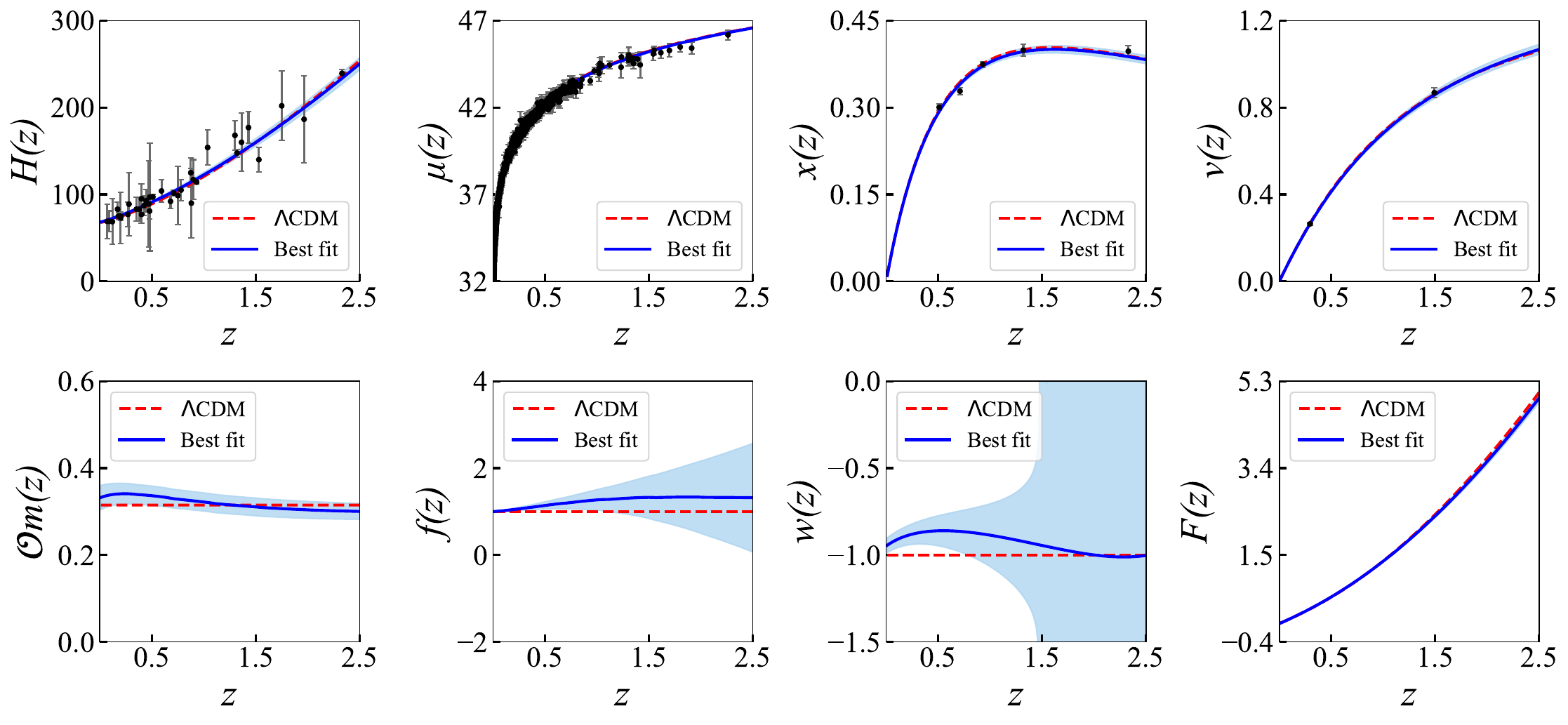}
\end{center}
\captionsetup{font=small} 
\caption{Marginalized posterior distribution of the set of semi-cosmographic parameters ($H_0$, $\Omega_m$, $\alpha$, $\beta$, $\gamma$) and the corresponding 2D confidence contours obtained from the MCMC analysis starting with $D_L^{\mathcal{R}}$. 
The confidence contours correspond to data being fitted with BAO (DESI DR1), CC,  Pantheon+, and joint (BAO + CC + Pantheon+), respectively.
The panel below shows the reconstruction of some diagnostics of background cosmology and their $1\sigma$ errors from the joint (BAO + CC + Pantheon+) analysis. 
The corresponding quantities for the $\Lambda$CDM model are  superposed for comparison.
}
\label{fig:reconstruct-R}
\end{figure*}

\begin{figure*}[htb]
\begin{center}
\includegraphics[height=7cm, width=7cm]{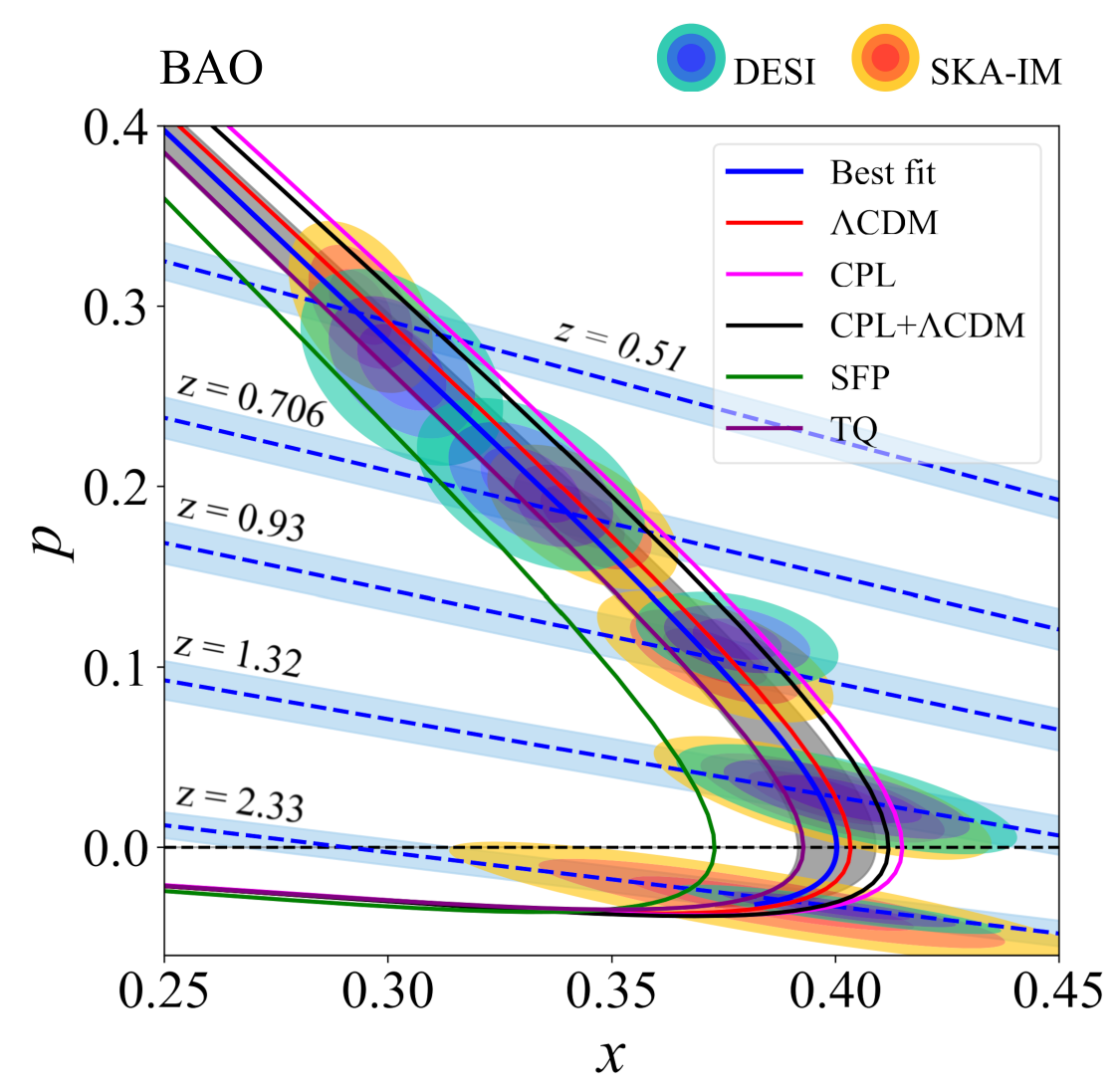}
\includegraphics[height=7cm, width=7cm]{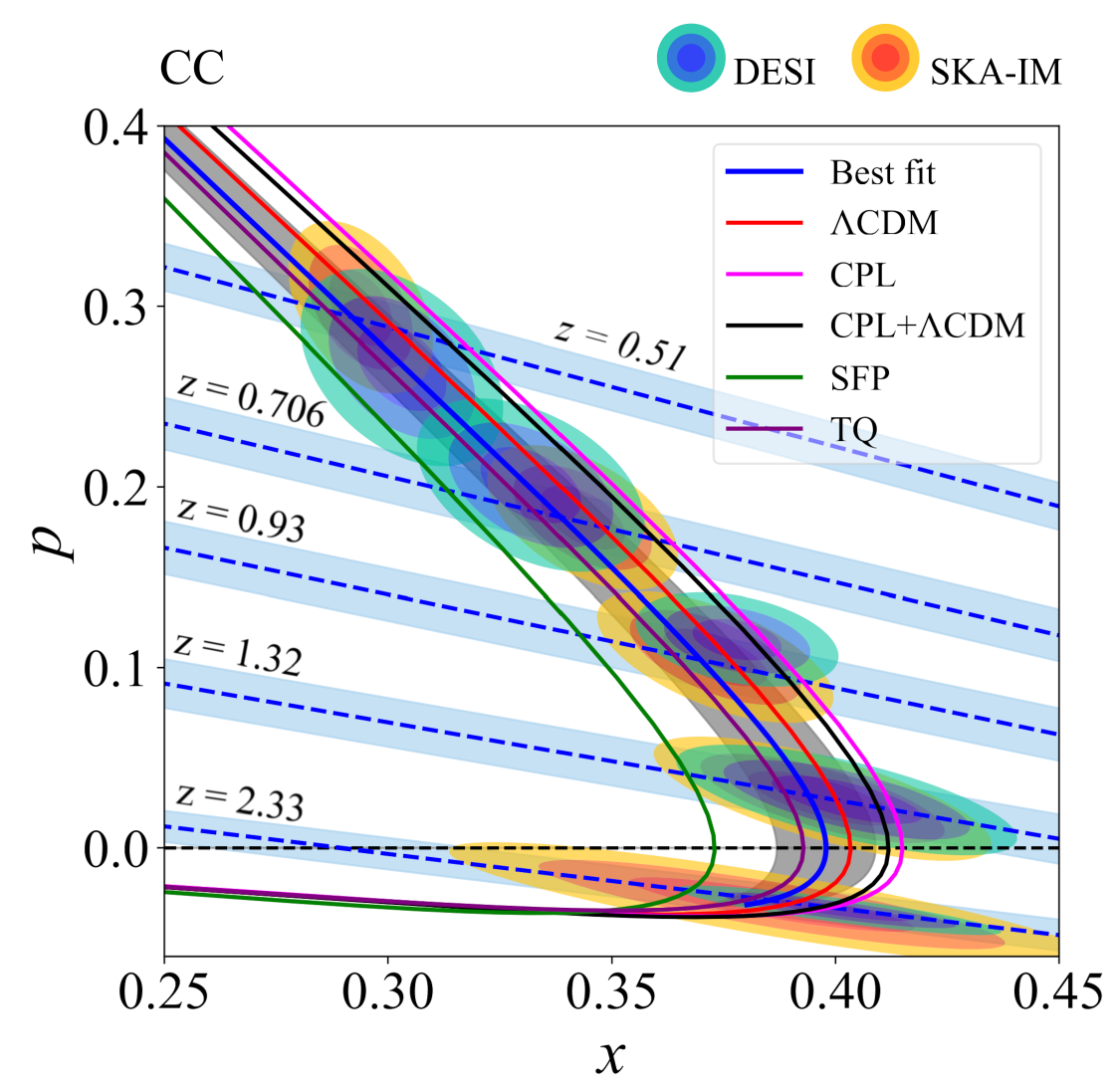}
\includegraphics[height=7cm, width=7cm]{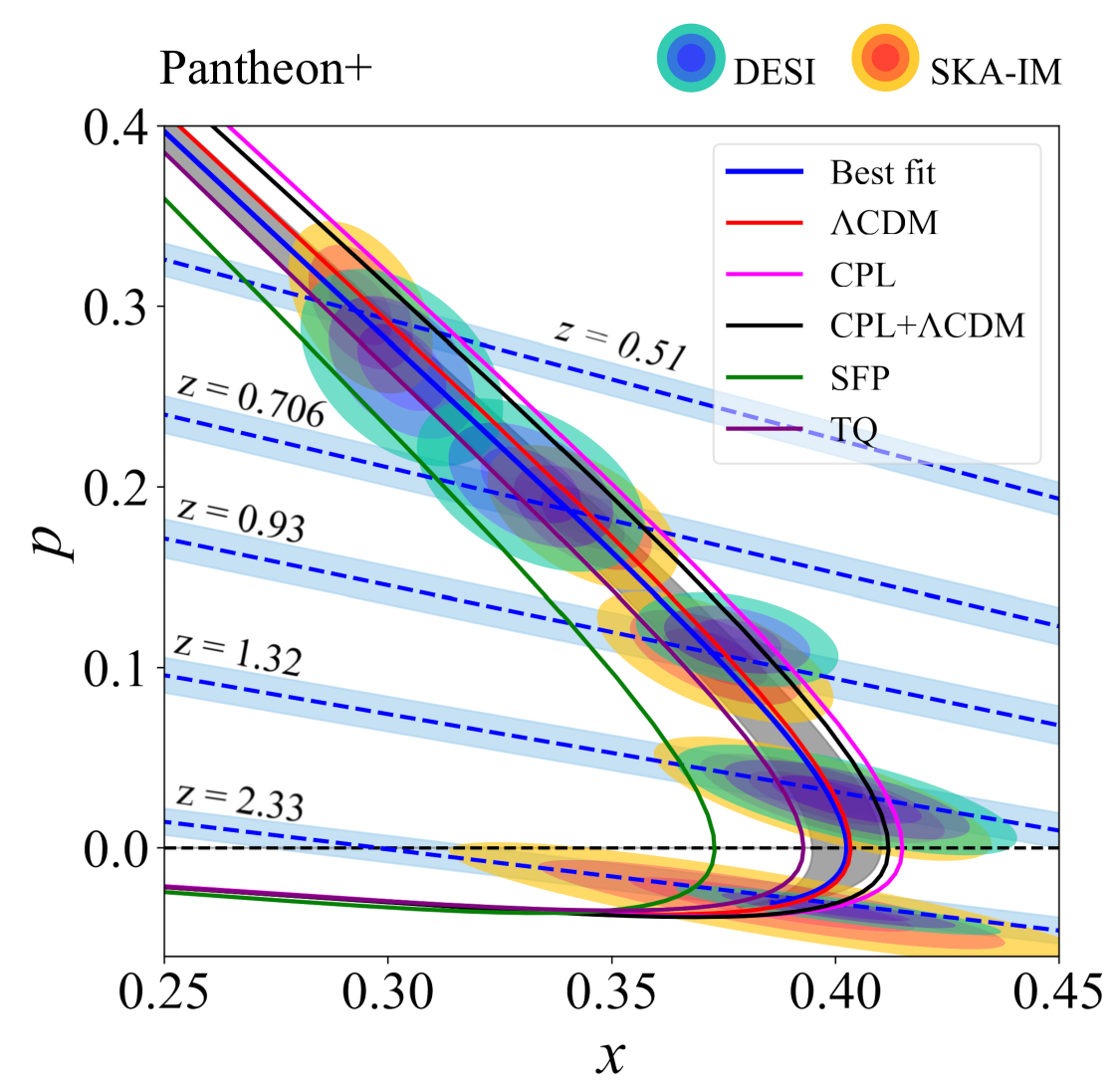}
\includegraphics[height=7cm, width=7cm]{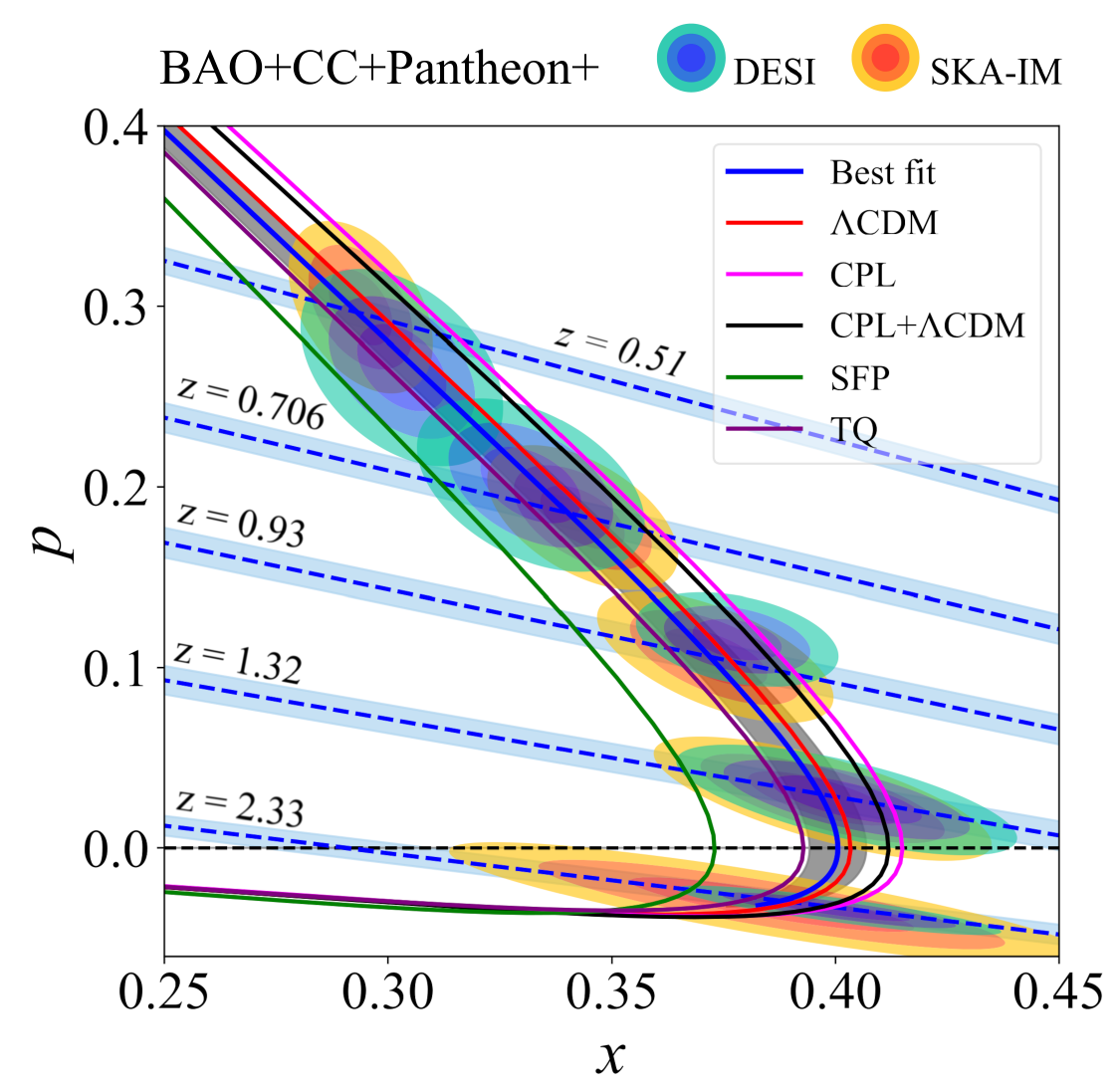}
\end{center}
\captionsetup{font=small} 
\caption{The reconstructed phase space trajectory $(x(z), p(z))$ with $1\sigma$ error for cosmography starting with $D_L^{{\cal R}}(z)$. Several DE models are also shown for comparison. 
The top-left figure corresponds to a reconstruction with BAO data, the top-right figure corresponds to CC data. The lower-left figure corresponds to a reconstruction using Pantheon+ data and the lower-right corresponds to a  joint analysis. 
The actual DESI error contours at 5 redshifts and the projected error contours for a 21-cm intensity mapping experiment at observing frequencies corresponding to the same redshifts are shown. We also show the consistency lines corresponding to the same redshifts with its error (originating from $H(z)$).}
\label{fig:phasespace-rec-R}
\end{figure*}

\section{Results and Discussion:}

We first discuss our results for the scenario described as $\bf Case ~I$. 
We consider the Pad\'e approximated luminosity distance of order $\left(2, 2\right)$ given by Eq.(\ref{eq:pade_P}). 
We choose $a_1=1$ so that we have
\be 
D_L^{\cal P} (z) = \frac{c}{H_0}\frac{z + a_2 z^2}{1 + b_1 z + b_2 z^2 }.
\label{eq:padeI} \ee
 The choice of $a_1 =1$ ensures that at very low redshifts the Pad\'e expansion and the Taylor expansion match, and all cosmological distances take a linear form ${cz}/{H_0}$.

This kinematic expansion and $D_A^{\cal P} $ and $H^{\cal P}$ obtained from Eq.(\ref{eq:padeI}) using Eq.({\ref{eq:pade1}}) has parameters $(H_0,  a_2, b_1, b_2)$. We obtain the semi-cosmographic equation of state
using Eq.(\ref{eq:eos}) which is then used to solve the dynamical system in Eq.({\ref{eq:autonomous}}) to obtain a new set $( D_A, H)$.
This new set,  now additionally depend on the parameter $\Omega_{m0}$. 
Figure \ref{fig:reconstruct-P} shows the results for fitting $(D_L, D_A, H)$ with BAO data from DESI, CC, SN data from Pantheon+ and  joint analysis with BAO + CC + Pantheon+.
The posterior distribution of fitting the same data  $D_A^{\cal P} $ and $H^{\cal P}$ are used as priors. The parameter estimation is done using MCMC (emcee code \cite{foreman2013emcee}). 
The best fit values of the parameters $ ( H_0, \Omega_{m0}, a_{1}, b_{1},  b_2)$ and the corresponding $1-\sigma$ errors are summarized in Table \ref{tab:MCMC-constraints}.
The value of $\chi^2_{red} \sim 1$, indicates that it is a good fit. 

The panel in figure \ref{fig:reconstruct-P} shows the best-fit reconstruction of ($H(z)$, $\mu(z)$, $x= H_0D_A(z)/c$, $v= D_V H_0/c$, ${\cal{O}}m(z)$, 
$f(z)$, $F(z)$, $w(z)$) with the $1\sigma$ errors obtained from the MCMC using  joint analysis with DESI DR1, CC,  and Pantheon+ data. The corresponding quantities for the $\Lambda$CDM model (with  Planck parameters) are also shown in the same figures for comparison. All the diagnostics seem to indicate that at $1\sigma$ the cosmographic model can not be distinguished from the $\Lambda$CDM model specially at high redshifts. The best fit for ${\cal{O}}m(z) > \Omega_{m0}$ seems to weakly favour quintessence models.
The worst constraint seems to be on $f(z)$ and the related equation of state parameter $w(z)$ which have large errors. 
$w(z)$ also seems to have an unphysical divergence at large redshifts. This is one of the key drawbacks of the cosmographic approach. The qualititative features of  the reconstructed diagnostics have similar behaviour as reported in literature for other model-independent/cosmographic approaches \cite{dutta2020beyond}.

Figure \ref{fig:phasespace-rec-P} shows the reconstructed phase space trajectory using BAO (DESI DR1), CC, Pantheon+ and joint (BAO + CC + Pantheon+) data respectively  for the semi-cosmographic analysis on 
$D_L^{\cal P}$. The best fit  phase trajectory and its $1\sigma$ error is shown. The lines of consistency are shown at the 5 redshifts corresponding to the DESI BAO data. The $1\sigma$ errors on these lines correspond to the uncertainties in the $p-$intercept which are related to the uncertainties in the reconstructed $H(z)$ at the specific $z$. The intersections of the 1-$\sigma$
band around the best fit trajectory with the bands around the lines of consistency gives the region of uncertainty of $(x,p)$ at a given redshift. We also show the phase space evolution for some cosmological models discussed in the earlier section.
We find that all these models are consistent with the best-fit result and indistinguishable within 1$\sigma$ for the analysis with CC data. The SFP model has tension with all the data sets.
The joint analysis with BAO, CC and Pantheon+ data, indicates that the best-fit cosmographic model is at a 2$\sigma$ tension with the CPL and CPL-$\Lambda$CDM  model and at almost a 3.5$\sigma$ tension with the SFP model.

The actual DESI DR1 data (transformed to the new variables) at the 5 redshifts are superposed on the reconstructed phase-space. At redshifts $z = 0.51$, the DESI results are  about $\sim 1.5 \sigma$
tension with the results of the reconstruction from the joint analysis.
We also show the projected 21-cm error contours at the same redshifts for a 21-cm  intensity mapping experiment described in the last section. We find that at low redshifts the 21-cm projections are competitive with the BAO DR1 results for an idealized (perfect foreground cleaning) intensity mapping experiment.

\renewcommand{\arraystretch}{1.25}
\begin{table}[h]
\small
\centering
\setlength{\tabcolsep}{2pt}
\begin{tabular}{l c c c c c c c}
\hline  \hline
 Model $D_L^{\cal P}$ &~~~~$H_0$ & $~~~~\Omega_{m0}$  & $a_2$ & $b_1$ &$b_2$& $\chi^2_{\rm red}$&AIC  \\ 
 \hline BAO & $69.48^{+1.2}_{-1.2}$ & $0.302^{+0.020}_{-0.026}$ & $1.343^{+0.099}_{-0.062}$ & $0.514^{+0.037}_{-0.033}$ & $-0.0001^{+0.0017}_{-0.0017}$& 1.997& 23.72\\ \hline
 CC & $68.49^{+2.8}_{-2.8}$ & $0.319^{+0.057}_{-0.044}$ & $1.340^{+0.160}_{-0.099}$ & $0.546^{+0.070}_{-0.079}$ & $-0.0001^{+0.0017}_{-0.0017}$& 0.561 & 25.14 \\ \hline
 Pantheon+ & $68.51^{+2.7}_{-2.7}$ & $0.303^{+0.043}_{-0.043}$ & $1.219^{+0.075}_{-0.092}$ & $0.467^{+0.051}_{-0.069}$ & $~~0.0000^{+0.0017}_{-0.0017}$ &0.886 & 1413.95\\ \hline
 BAO+CC+Pantheon+ & $68.06^{+0.42}_{-0.42}$ & $0.306^{+0.016}_{-0.023}$ & $1.226^{+0.031}_{-0.031}$ & $0.467^{+0.018}_{-0.018}$ & $~~0.0002^{+0.0015}_{-0.0015}$ &0.879 & 1441.80\\ 
 \hline \hline \\

 \hline  \hline
 Model $D_L^{\cal P}$ &$H_0$ & $\Omega_{m0}$  & $\alpha$ & $\beta$ &$\gamma$& $\chi^2_{\rm red}$&AIC \\ 
 \hline BAO & $68.05^{+1.0}_{-1.0}$ & $0.301^{+0.047}_{-0.042}$ & $1.295^{+0.16}_{-0.13}$ & $0.637^{+0.026}_{-0.026}$ & $-0.282^{+0.048}_{-0.048}$& 2.848& 29.94\\ \hline
 CC & $67.94^{+2.6}_{-2.6}$ & $0.302^{+0.035}_{-0.066}$ & $1.322^{+0.20}_{-0.16}$ & $0.626^{+0.032}_{-0.032}$ & $-0.245^{+0.054}_{-0.054}$& 0.548 & 24.79 \\ \hline
 Pantheon+ & $68.60^{+2.7}_{-2.7}$ & $0.302^{+0.054}_{-0.060}$ & $1.190^{+0.17}_{-0.17}$ & $0.627^{+0.023}_{-0.023}$ & $-0.239^{+0.047}_{-0.047}$ &0.886 & 1414.19\\ \hline
 BAO+CC+Pantheon+ & $67.94^{+0.42}_{-0.42}$ & $0.295^{+0.013}_{-0.030}$ & $1.250^{+0.12}_{-0.10}$ & $0.647^{+0.019}_{-0.019}$ & $-0.297^{+0.037}_{-0.037}$ &0.879 & 1442.62\\ 
 \hline \hline \\
\end{tabular}
\captionsetup{font=small} 
\caption{The parameter values obtained in the MCMC analysis are tabulated along with the $1\sigma$ uncertainty.}
\label{tab:kinematics}
\label{tab:MCMC-constraints}
\end{table}

\renewcommand{\arraystretch}{1.25}
\begin{table}[h]
\small
\centering
\begin{tabular}{l c c c }
\hline  \hline
 Model $D_L^{\cal P}$ &$q_0$ & ${j_0}$  & $s_0$  \\  %[0.1ex] 
 \hline   
 DESI &$-0.610^{+0.033}_{-0.076}$& $2.30^{+0.69}_{-0.22}$  & $5.6^{+2.6}_{-1.0}$   \\ \hline
 CC &$-0.532^{+0.078}_{-0.130}$& $1.93^{+1.00}_{-0.39}$  & $4.3^{+2.3}_{-2.3}$  \\ \hline
 Pantheon+ &$-0.526^{+0.055}_{-0.066}$& $1.62^{+0.60}_{-0.60}$  & $3.3^{+1.2}_{-1.0}$ \\ \hline
 BAO+CC+Pantheon+ &$-0.540^{+0.037}_{-0.037}$& $1.66^{+0.25}_{-0.32}$  & $3.3^{+0.7}_{-1.0}$\\
 \hline \hline \\
\end{tabular}
\captionsetup{font=small} 
\caption{The best  fit values of  $(q_0, j_0, s_0)$  along with the corresponding $1\sigma$ errors for a Pad\'e cosmography with $D_L^{\cal P}$.}
\label{tab:kinematics}
\end{table}

In our analysis, we have made no assumption about the connection between a Pad\'e approximation and Taylor series expansion. 
However, for the form of $D_L^{\cal P}(z)$ chosen by us, such a comparison is possible and the parameters $a_2$, $b_1$ and $b_2$ can be expressed in terms of the kinematic quantities 
$q_0$, $j_0$ and $s_0$. Using the relationship from \cite{Capozziello_2020_cosmography, Capozziello_2018_cosmography},  we obtain the constraints on these kinematic quantities.
Table \ref{tab:kinematics}summarizes the constraints on $(q_0, j_0, s_0) $, if these parameters were used in $D_L^{{\cal P}}$ instead of $(a_2, b_1,b_2)$.
The best fit values of these kinematic parameters are consistent with the findings in other cosmographic methods \cite{Capozziello_2018_cosmography}.

We shall now  discuss the scenario described as $\bf Case ~II$. 
Here we have  the Pad\'e approximated luminosity distance in the variable $(1+z)^{1/2}$ instead of $z$ given by Eq.(\ref{eq:pade_R}).
For this model $ H(z) \rightarrow H_0$ as $z \rightarrow 0$ and $H(z) \propto (1+ z)^{3/2}$ for $z>>1$.
We first fit $D_L^{\cal R}(z)$ and $D_A^{\cal R}$ and $H^{\cal R}$ using parameters $(H_0, \alpha, \beta, \gamma)$ with data using flat priors.
The posteriors from these fits are then used as priors for fitting the semi-cosmographic $D_L(z)$, $D_A(z)$ and $H(z)$ obtained by solving Eq.(\ref{eq:autonomous}) using $w^{[ \Omega_m, \cal R]} (z)$. We take flat priors for $\Omega_{m0}$ as before.

The estimated parameters and their $1\sigma$ errors are given in Table \ref{tab:MCMC-constraints}.
The constraints on $H_0$ and $\Omega_{m0}$ are comparable to the ones obtained using $D_L^{P}(z)$.

 Figure \ref{fig:reconstruct-R} shows the results for fitting $(D_L, D_A, H)$ with BAO data (DESI DR1), CC data, SNIa data from Pantheon+ and  joint analysis with BAO + CC + Pantheon+. The reduced $\chi^2 \sim 1$ implying that the fit is good.

The panel in figure \ref{fig:reconstruct-R} shows the best-fit reconstruction of ($H(z)$, $\mu(z)$, $x= H_0 D_A(z)/c$, $v= D_V H_0/c$, ${\cal{O}}m(z)$, 
$f(z)$, $F(z)$, $w(z)$) with the $1\sigma$ errors obtained from joint analysis with BAO (DESI DR1), CC and Pantheon+ data. The corresponding quantities for the $\Lambda$CDM model (with Planck parameters) are also shown in the same figures for comparison. Here too, all the diagnostics show that the semi-cosmographic model $\cal R$ can't be distinguished from the $\Lambda$CDM model. At low redshifts the departure is $ \sim 1 \sigma$.
In this case however $w(z)$ undergoes a pathological divergence at $\sim z >  2.5$. This makes the DE equation of state a poorly constrained function with little information about cosmic evolution  at large redshifts. The qualitative features of the other reconstructed diagnostics have similar behaviour as those obtained from $D_L^{\cal P}$. 
 We have calculated the AIC (Akaike Information Criterion) \cite{Akaike_1974_AIC} to test  which of the two cosmographic models perform better towards fitting parameters with data. We find that for all the data sets, there is not much  difference in the AIC, which indicates that there is no favourable choice out of the two cosmographic scenarios.  

Figure \ref{fig:phasespace-rec-R} shows the reconstructed phase-space trajectory for $D_L^{\cal R}$ as the starting point of the semi-cosmographic analysis. 
 While $\Lambda$CDM  model is  consistent with the reconstructed phase trajectory, the CPL model and the CPL-$\Lambda$CDM  model (with their model parameters  obtained by fitting with other data sets) are at a $2\sigma$ tension with our reconstructed result. The SFP model has a $\sim 3 \sigma$ tension with the best fit result. The low redshift DESI results also seem to have a weak tension $\sim 1.5 \sigma$ with the reconstructed estimates using the joint analysis.

\section{Summary and Conclusion}
We have developed a description of cosmological evolution in the phase space of dimensionless variables $x = H_0 D_A/c$ and $p = dx/dz$.  
This phase space approach focuses our attention to the fact that $H(z)$ and $D_A(z)$  can be independently measured at a  given redshift, which  allows us to study them simultaneously, instead of seeing them separately as functions of $z$.
In the standard cosmography $H(z)$ and $D_A(z)$ are reconstructed as a function of $z$. 
Each of these evolutions carry half the information (since the  dynamical  Friedmann equation is a second order differential equation). Thus, it is meaningful to study them together in a phase space by eliminating $z$ between $H(z)$ and $D_A(z)$. Since  $H(z)$ and $D_A(z)$ are independently measured, it is required to study both $H(z)$ and $D_A(z)$.
When  $(D_A, H)$  is studied in the phase space in our equivalent  approach, it gives a geometrical (phase space) interpretation of these two cosmological quantities of interest.
All possible theoretical models are curves in the accessible region of the phase space which must merge at $z=0$ and $z=\infty$.
Any observational data at any redshift  will be a point in the phase space with a region of uncertainty around it. Such an observational data can, thus  be directly compared with theoretical models (which are curves in the phase space).
Thus, in our proposed method, the compatibility of any observational data with any theoretical model is directly tested, instead of checking  them  separately for $H(z)$ and $D_A(z)$.
Using many data points, one may reconstruct the best fit phase space trajectory, giving a robust method to rule out theoretical models.

To integrate the dynamical system $(x(z),p(z))$ we have refrained from showing any preference for specific DE models. We consider two kinematic models where the Luminosity distance is expanded as Pad\'e rational approximants using expansion in terms of $z$ and $(1+z)^{1/2}$ respectively and  solved the dynamical problem in the phase space by constructing a semi-cosmographic equation of state for DE. The semi-cosmographic $(D_L(z), D_A(z), H(z))$, thus obtained are fitted with BAO and SNIa data from DESI DR1 and Pantheon+ respectively. We have also used CC data in the analysis. Further, we  have also  considered projected error covariances for a  futuristic SKA like 21-cm intensity mapping experiment. In the absence of foregrounds the error projection from the 21-cm intensity mapping is competitive with DESI DR1. However, strong foreground residuals shall degrade these projections significantly. 
We have assumed naively the difference in the spectral properties of the signal from those of the foreground has been used for complete foreground cleaning \cite{2011MNRAS.418.2584G, Ghosh_2010, liu2009improved, liu2012well, wang200621}.
The foregrounds from Galactic synchotron emission and extragalactic point sources are several orders larger than the signal. However, the foregrounds are spectrally smooth and thus, in principle contaminate very small line-of-sight
wave modes. Modeling and subtracting a low order polynomial is typically to be performed \cite{Ghosh_2010}.  In reality  for a  3D power spectrum estimation using a visibility-visibility correlation approach, spectrally smooth foregrounds contaminate the {\it foreground wedge} in the $(k_{\parallel}, k_{\perp})$ space due to the frequency dependence (chromatic)  of the interferometer's fringe pattern.  While a perfect knowledge of the telescope resposne can in principle allow us to clean the
foregrounds, one may leave the modes in the wedge and use the clean window in $k-$ space. For a  BAO observation this leads to a significant degradation. It has been studied that the  $z \sim 1 -2$ the wedge effect causes  the errors on $D_A$ to be
increased by $3$ to $4.4$ times. 
The errors on $H(z)$ may be enhanced by a factor $\sim 1.5$ at these redshifts. \cite{Seo_2016_wedge}. In our work we have assumed perfect foreground cleaning and not incorporated the wedge effect.

Further, we have  used the semi-cosmographic fitting to reconstruct some diagnostics of background cosmology and compared our results for the two scenarios of Pad\'e expansions. 
The reconstructed diagnostics point towards dynamical DE. The equation of state reconstructed in a cosmographic manner has divergences and are not well behaved in the entire parameter space.
There are two issues here. Firstly, we note that 
any model independent approach to reconstruct $w(z)$ will have this problem. The denominator in the expression for $w(z)$ will approach zero when $(H/H_0)^2$ approaches $\Omega_{m0} (1+z)^3$. One way to push this divergence to higher redshifts is to include  radiation always. 
Obviously at low redshifts radiation will not play any role. But at high redshifts will help to avoid the denominator from going to zero.
One may also numerically impose hard priors in the MCMC analysis or impose high cost in the likelihood to avoid such divergences. 

The second issue is that in cosmography we are starting with Luminosity distance and then arrive at $w(z)$ after twice differentiate the $D_L(z)$.
Each differentiation makes the error bars larger as we are differentiating a noisy data. This makes the error bar on $w(z)$ very large. This gets worse at  high redshifts where there is hardly  any data and hence $D_{L}$ is poorly reconstructed at high redshifts.  Since we are differentiating this highly unknown $D_{L}$ to get $w(z)$, it makes the error bars of $w(z)$ to blow up.
This actually indicates  that $w(z)$ is not a good diagnostic specifically at high redshifts as we do not have enough data to constraint  $D_{L}$ or $D_{A}$ meaningfully.

This makes the semi-cosmographic parameter estimation challenging. However, since we are solving a system of differential equations, error accumulation through a double integration to go from $w(z)$ to $H(z)$ to $D_A(z)$ is avoided. There is nothing special about $D_L$ being the starting observable expanded in a Pad\'e series. It could have been any other distance or even the Hubble parameter. The method shall go through in the same way. We conclude by noting that the cosmological evolution in phase space shall get better constrained with future data from precision observations.

\acknowledgments
CBV acknowledges the National Research Foundation (NRF) Postdoctoral Fellowship, South Africa, for financial support. AAS acknowledges the funding
from ANRF, Govt of India, under the research grant no.
CRG/2023/003984.
\bibliographystyle{JHEP}
\bibliography{references}
\end{document}